\documentclass[proceedings]{JHEP3}
\usepackage{eurosym}
\usepackage{amssymb}
\usepackage{braket}
\usepackage{amsfonts}
\usepackage{amsmath,epsfig}
\usepackage{graphicx}
\usepackage{multirow}
\usepackage{subcaption}
\usepackage{tikz}
\usepackage{float}
\usepackage{xcolor}
\newbox\mybox

\newcommand\fverb{\setbox\mybox=\hbox\bgroup\verb}
\newcommand\fverbdo{\egroup\medskip\noindent\fbox{\unhbox\mybox}\ }
\newcommand\fverbit{\egroup\item[\fbox{\unhbox\mybox}]}
\conference{Non-Hermitian gauge field theories and BPS limits}
\abstract{We present an overview of some key results obtained in a recent series devoted to non-Hermitian quantum field theories for which we systematically modify the underlying symmetries. Particular attention is placed on the interplay between the continuous symmetry group that we alter from global to local, from Abelian to non-Abelian, from rank one to generic rank N, and the discrete anti-linear modified CPT-symmetries. The presence of the latter guarantees the reality of the mass spectrum
in a certain parameter regime. We investigate the extension of Goldstone's theorem and the Higgs mechanism, which we demonstrate to work in the conventional fashion in the CPT-symmetric regime, but which needs to be modified technically at the standard exceptional points of the mass spectrum and entirely fails at what we refer to as zero exceptional points as well as in the broken CPT-symmetric regime. In the full non-Hermitian non-Abelian gauge theory we identify the t'Hooft-Polyakov monopoles by means of a fourfold Bogomol'nyi-Prasad-Sommerfield (BPS) limit. We investigate this limit further for other types of non-Hermitian field theories in 1+1 dimensions that possess complex super-exponential and inverse hyperbolic kink/anti-kink solutions and for 3+1 dimensional Skyrme models for which we find new types of complex solutions, that all have real energies due to the presence of different types of CPT-symmetries.}

\title{Non-Hermitian gauge field theories and BPS limits}
\author{Andreas Fring and Takanobu Taira \\
Department of Mathematics, City, University of London,\\
Northampton Square, London EC1V 0HB, UK \\
E-mail: a.fring@city.ac.uk, takanobu.taira@city.ac.uk}

\input{tcilatex}
\begin{document}

\section{Introduction}

Abelian and non-Abelian gauge field theories as manifested in quantum electrodynamics and the standard model, respectively, play a central role in particle physics and quantum field theory. While the general principle of 
gauge theories is a powerful universal concept, the concrete models have their well-known limitations see e.g. 
\cite{ellis2002limits,ellis2012outstanding} and the very recent evidence for the breaking of lepton universality in beauty-quark decays \cite{lepuni}. Many extensions of the standard model by new types of ideas such as supersymmetry or the extension to string/M-theory have and are being explored. Here we report on a series of studies \cite{FringGoldstone,FringGoldstone2,FringMonopole,FringHiggs,FringBPS,FringSkyrmion} that explore certain sectors of the standard model to allow for the possibility of non-Hermitian gauge field theories.

This approach in field theory is inspired by the success of non-Hermitian extensions to quantum mechanics. In the conventional approach to quantum mechanics one usually demands Hamiltonians to be Hermitian, as this property guarantees the reality of the energy spectrum and unitary time-evolution. However, in a seminal paper Bender and Boettcher \cite{bender_boettcher_real_spectrum} realised that Hermiticity is only a sufficient but not a necessary condition to ensure these properties. The reality of the spectrum may still be guaranteed when the Hamiltonian and the corresponding wave functions respect an anti-linear symmetry \cite{EW} of which a simultaneous parity and time-reversal ($\mathcal{PT}$)-symmetry is an example. When modifying the inner product by a suitable new metric, also the unitary time-evolution, i.e. the preservation of probability densities, can be ensured. Meanwhile these idea have found their way into a wide range of areas in physics with classical optics being especially successful as in that context many experiments and applications can be realised, see e.g. \cite{PT_optics_ref_1,PT_optics_ref_2,PT_optics_3,PT_optics_4,PT_optics_5}, thus not only confirming the theoretical formulation but also supporting the manifestation of  $\mathcal{PT}$ symmetric systems in nature. 
    
    The natural extension of these ideas to the quantum field theories is still far less developed, but is currently an active field of research. In this review we shall mainly focus on the study of various classical solutions to field theories. We start by investigating how the central concepts in standard Hermitian quantum field theory, such as how the Higgs mechanism builds on the Goldstone theorem, are extended to non-Hermitian theories. We recall that the former
    predicts the number of massless Nambu-Goldstone fields \cite{goldstone1961field,nambu1961dynamical} as the number of global continuous symmetry generators of the theory that are broken by the vacuum around which the theory is expanded. For a concrete theory with global $U(1)$ and $SU(N)$-symmetry with two or three fields in the adjoint and fundamental representation we discuss how the discrete anti-linear symmetries determine a parameter regime in a non-Hermitian theory in which the theorem still holds, needs to be technically modified and eventually completely breaks down. When changing the global to a local symmetry the Higgs mechanism \cite{Higgs_mechanism_ref_1,Higgs_mechanism_ref_2,Higgs_mechanism_ref_3,Higgs_mechanism_ref_4} couples these fields to gauge fields in such a way that the combined field acquires a mass, such as for instance the W and Z gauge bosons in the standard model. 
    
    The non-Hermitian gauge theories obtained in this manner are then further probed for their existence of magnetic monopoles that constitute another important type of classical solutions of the quantum field theory. For our non-Hermitian toy model these solutions are obtained in a fourfold scaling limit of the corresponding equation of motion usually referred to as the 
    Bogomol'nyi-Prasad-Sommerfield (BPS) limit \cite{Bogomolny_scaling_limit,Prasad_Sommerfield_scaling_limit}. These type of solutions are special as they saturate the lowest energy or Bogomol'nyi bound. Here we pay particular attention to whether these energies are still real despite solving a complex equation. We provide a reality condition whose key ingredient is a modified $\mathcal{CPT}$-symmetry of the Hamiltonian and the solutions. 
    
    Subsequently we adopt the viewpoint to treat these type of non-Hermitian BPS theories in their own right and investigate them in 1+1 and also 3+1 dimensions in form of reduced versions of the Skyrme model \cite{Skyrme_original} that was designed to model the nuclei. 
    
    Our manuscript is organised as follows. In section 2 we elaborate on the spontaneous symmetry breaking in non-Hermitian field theories with a discussion of Goldstone's theorem in section 2.1 and the Higgs mechanism in section 2.2. Complex solutions to BPS equations are discussed in section 3, with t'Hooft-Polyakov magnetic monopoles in section 3.1, dual BPS theories in 1+1 dimensions and BPS Skyrmion solutions in section 3.3. Our conclusions are stated in section 4.
    
\section{Spontaneous symmetry breaking in non-Hermitian field theory}\label{section: SSB in non-Hermitian theory}
In the context of non-Hermitian $\mathcal{PT}$ symmetric quantum mechanics \cite{PTbook} it is well known that one needs to find a well-defined metric to ensure a unitary time-evolution of a Hamiltonian system and also to define meaningful observable quantities. A direct investigation of a non-Hermitian system therefore often leads to bizarre conclusions and apparent inconsistencies, as for instance when studying properties of non-observable quantities in a conventional framework with an inconsistent inner product. These issues persist and carry over to a quantum field theoretic setting, where a similar problem arises for instance when one studies the variational principle for an action of a complex scalar field theory $I=\int \mathcal{L}(\phi,\phi^*)$. The equations of motion obtained from  $\delta I/ \delta \phi =0$ and $\delta I / \delta \phi^* =0$ appear to be incompatible under complex conjugation due to the fact that $I\not= I^*$. One may attempt to remedy these issues by all kind of mechanism, e.g. \cite{AlexandreNoether,Alexandre_Noether_2}, but here we adopt the well-established and successful viewpoint relying on the construction of what is often referred to as the pseudo-Hermitian approach \cite{Alirev,PTbook} by finding a Dyson map $\eta$ \cite{Dyson}, whose adjoint action map a non-Hermitian Hamiltonian $H$ to a Hermitian Hamiltonian $h$ as $\eta H\eta^{-1} = h$.

Technically these transformations are usually difficult to find and are also known to be not unique \cite{Urubu}. Here we encounter two versions of these transformations, as a map acting on the field theoretic Hamiltonian involving non-commutative equal time commutation relations between canonical fields and also as matrix transformations on the squared-mass matrix. We will see that for some examples the action of the field of these maps is indeed identical.
For references purposes and in order to establish our notation we briefly recall the standard technique used for finite dimensional Hamiltonians $H$, where in the low order field theoretic context the non-Hermitian squared-mass matrix $M^2$ is the analogue to $H$.

As a starting point the Hamiltonian is assumed to be $\mathcal{PT}$-symmetric, that is a simultaneous parity $\mathcal{P}$ and time transformation $\mathcal{T}$,
\begin{equation}\label{properties of P}
    [H,\mathcal{PT}]=H\mathcal{P}-\mathcal{P}H^*=0 ~,~~~ \mathcal{P}^T \mathcal{P}=\mathbb{I}.
\end{equation}
A bi-orthonormal basis can then be constructed from the right and left eigenvector, $\{v_n\}$ and $\{u_n\}$, respectively, of $H$ 
\begin{equation}
    Hv_n = \epsilon_n v_n ~,~~~ H^\dagger u_n = \epsilon u_n,
\end{equation}
where the dagger denotes the usual complex conjugate transpose. This basis satisfies the following orthonormality and completeness relations 
\begin{equation}\label{properties of bi-orthonormal basis}
    \braket{u_n |v_m} =\delta_{nm} ~,~~~  \sum_n \ket{u_n}\bra{v_n} = \sum_n \ket{v_n}\bra{u_n} = \mathbb{I},
\end{equation}
with the left and right eigenbasis related to each other by the parity operator 
\begin{equation}\label{relation between left and right bi-orthonormal basis}
    \ket{u_n} = s_n \mathcal{P} \ket{v_n},
\end{equation}
with $s_n = \pm 1$ defining the signature. Combining the relations (\ref{properties of bi-orthonormal basis}), (\ref{relation between left and right bi-orthonormal basis}) and the second relation in (\ref{properties of P}) we can express the parity operator as 
\begin{equation}\label{definition of P}
    \mathcal{P} = \sum_n s_n \ket{u_n}\bra{u_n} ~,~~~ \mathcal{P}^T =\sum_n s_n \ket{v_n}\bra{v_n}.
\end{equation}
The relation $\mathcal{P}^T \mathcal{P}=1$ then automatically holds, following from $\mathcal{P}^T \mathcal{P}\ket{v}=\ket{v}$ and $\bra{u}\mathcal{P}^T \mathcal{P}=\bra{u}$. 

Closely related, one defines an operator $\mathcal{C}$ as
\begin{equation} \label{definition of C}
    \mathcal{C} =  \sum_n s_n \ket{v_n} \bra{u_n}, 
\end{equation}
that is easily seen to be related to the metric operator $\rho: = \eta^\dagger \eta $ as $ \mathcal{C} = \rho^{-1}\mathcal{P}$. The operator $\mathcal{C}$ satisfies the following algebraic relations
\begin{equation}\label{three algebraic conditions for C}
    [\mathcal{C},H]=0~,~~~ [\mathcal{C},\mathcal{PT}]=0~,~~~ \mathcal{C}^2=\mathbb{I}.
\end{equation}
Naturally when identifying the $\mathcal{C}$ and $\mathcal{P}$ operators at the level of the squared-mass matrix we have ignored all interaction terms, which might not be left invariant. We will see below that requiring $\mathcal{PT}$ to be a symmetry of the entire Lagrangian will select out a particular subset, which for our solutions restricts the possible choices of the signatures $\{s_n \}$.  
    \subsection{Nambu-Goldstone bosons in models with global $U(1)$ and $SU(2)$ symmetry}
    We begin by considering a theory that respects a continuous global $SU(N)$ symmetry and a discrete modified anti-linear $\mathcal{CPT}$-symmetry. The following model is a direct generalization of a $U(1)$ symmetric model first studied in the context non-Hermitian field theories in \cite{AlexandreGoldtone}
        \begin{eqnarray}
            \mathcal{L}_n^{SU(N)} = \sum_{i=1}^n \partial_\mu \phi_i^\dagger \partial^\mu \phi_i + c_i m_i^2 \phi_i^\dagger \phi_i +\sum_{i=1}^{n-1}\kappa_i \mu_i^2 \left(\phi_i^\dagger \phi_{i+1}-\phi^\dagger_{i+1}\phi_i\right) - \frac{g_i}{4}\left(\phi_1^\dagger \phi_1\right)^2,
        \end{eqnarray}
       containing $n$ complex scalar fields $\phi_i = (\phi_i^1, \dots , \phi_i^N)$, $i, \dots n$, with $N$-components, where each field $\phi_i$ is taken to be in the fundamental representation of $SU(N)$. The constant parameters in the model are $m_i ,\mu_i , g_i \in \mathbb{R}$ and $\kappa_i ,c_i \in \{-1,1\}$. The respective continuous $SU(N)$ and discrete anti-linear $\cal {CPT}$ symmetries are realised as \cite{FringGoldstone}
       \begin{eqnarray} \label{symm}
       	SU(N) &:&\phi _{j}\rightarrow e^{i\alpha T^{a}}\phi _{j},~\ \ \ ~\ \  \\
       	{\cal CPT}_{1/2} &:&\phi _{i}(x_{\mu })\rightarrow \mp \phi _{i}^{\ast
       	}(-x_{\mu })~~\text{for }\frac{i}{2}\in {\Bbb Z},~~\phi _{j}(x_{\mu
       	})\rightarrow \pm \phi _{j}^{\ast }(-x_{\mu })~~\text{for }\frac{j+1}{2}\in 
       	{\Bbb Z}, \qquad
       \end{eqnarray}%
       with real parameters $\alpha $ and $SU(N)$ generators $T^{a}$. Notice that the $\cal {CPT}$ symmetry is not the standard one commonly assumed in Hermitian theories.
   
        A key ingredient in the pseudo-Hermitian approach is to find an equivalent Hermitian Lagrangian to the original non-Hermitian one by acting on it adjointly with the Dyson map. In our case we can use the map 
        \begin{equation}
            \eta = \exp\left[\frac{\pi}{2}\sum_{i=2,4,6,\dots}\sum_{\alpha=1}^N \int d^3 x \Pi^{\varphi_i^\alpha} (t,\Vec{x})\varphi_i^\alpha (t,\Vec{x})+\Pi^{\chi_i^\alpha}(t,\Vec{x})\chi_i^\alpha (t,\Vec{x}) \right], \label{DM}
        \end{equation}
        expressed in terms of the real components $\varphi_i^\alpha$, $\chi_i^\alpha$ of the complex fields $\phi_i^\alpha = (\varphi_i^\alpha + i \chi_i^\alpha)/\sqrt{2}$ to transform , $\mathcal{L}_n^{SU(N)}$. This Dyson map is a generalisation of the one used in \cite{MannheimAntilinear,MannheimHiggs}, see $S$ after equations (37) and (40), respectively, and is, as is well-known, not unique. It does not exhibit the standard features of becoming ill-defined at the exceptional points of the theory and in the broken $\mathcal{PT}$-regime in an obvious manner. Furthermore, in 
        \cite{Alexandre_Ref_on_ill_defined_Dyson_map} it has been reported that issues with ghost fields arise possibly due to the application of this map. Here we will proceed and use (\ref{DM}) to construct an equivalent Hermitian system.
        For this purpose we resort temporarily to the quantum theory and when assuming the standard equal-time commutation relation between the fields and their corresponding conjugate momentum operators, each of the real component fields transforms as 
        \begin{equation}
            \varphi_{2i}^\alpha \rightarrow -i \varphi_{2i}^\alpha ~,~~~ \chi_{2i}^\alpha \rightarrow -i \chi_{2i}^\alpha .
        \end{equation}
        When recasting the transformed equivalent Hermitian Lagrangian into a simpler form by defining the $N\times n$-component vector field $\Phi=(\varphi_1^1,\chi_2^1,...,)$ we obtain 
        \begin{equation}
            \mathcal{L} =\frac{1}{2} \partial_\mu  \Phi \mathcal{I} \partial^\mu \Phi + \frac{1}{2} \Phi^T H \Phi + \mathcal{O} (\Phi^4) .
        \end{equation} 
        Notice that as a result of the transformation we have introduced the metric operator $\mathcal{I}=\text{diag}(1,-1,1,\dots)$ in the kinetic term. Crucially the Hessian matrix $H$ is now Hermitian.  The squared-mass matrices of the above Lagrangian are then found by calculating the expression $M^2:=\mathcal{I}H$. 
        
        Depending on which vacuum we choose to expand around, we expect to find different types of spectra containing the number of massless Goldstone bosons as predicted by Goldstone's theorem \cite{goldstone1961field,nambu1961dynamical}, i.e. when expanding around the $SU(N)$-symmetric vacua we expect to find no massless Goldstone boson, whereas when expanding around the vacua that break the $SU(N)$-symmetry we anticipate to have a massless Goldstone boson in the spectrum for each broken generator of the symmetry group. In \cite{FringGoldstone} we confirmed that these standard predictions from Hermitian theories also hold in the regions of the parameter space where the $\mathcal{CPT}$ is unbroken and at the boundaries of those domains. However, besides predicting the right number of zero eigenvalues for the theory to be meaningful, one also needs to be able to identify the fields corresponding to the Goldstone bosons. It is this latter property that breaks down at the boundaries of the $\mathcal{CPT}$-symmetric regime. Let us see in some more detail what kind of boundaries and physical regions we may encounter.
        
        For this purpose we need to compute the squared-mass matrices by expanding around specific vacua. For instance, for the Abelian $U(1)$-symmetric model with $n=3$ fields and the non-Abelian $SU(2)$-model with $n=2$ fields expanded around the symmetry broken vacua we obtain the squared-mass matrices
        \begin{equation}
            M^{2}_{U(1),3}=\left(\begin{array}{cccccc}
                 \frac{3c_3 m_3^2 \mu^4}{c_2 c_3 m_2^2 m_3^2 +\nu^4}+2c_1 m_1^2 & -c_\mu \mu^2&0&0&0&0  \\
                 c_\mu \mu^2&-c_2 m_2^2 &c_\nu \nu^2&0&0&0\\
                 0&-c_\nu \nu^2 &-c_3 m_3^2&0&0&0\\
                 0&0 &0&\frac{c_3 m_3^2\mu^4}{c_2 c_3 m_2^2 m_3^2 +\nu^4}&c_\mu \mu^2&0\\
                 0&0 &0&-c_\mu \mu^2&-c_2 m_2^2&-c_\nu \nu^2\\
                 0&0 &0&0&c_\nu \nu^2&-c_3 m_3^2\\
            \end{array}\right),
        \end{equation}
        \begin{equation}\label{Global SU(2) mass matrix}
            M_{SU(2),2}^{2}= \left(
                \begin{array}{cccccccc}
                 -\frac{\mu ^4}{m_2^2} & \mu ^2 & 0 & 0 & 0 & 0 & 0 & 0 \\
                 \mu ^2 & -m_2^2 & 0 & 0 & 0 & 0 & 0 & 0 \\
                 0 & 0 & \frac{\mu ^4}{m_2^2} & -\mu ^2 & 0 & 0 & 0 & 0 \\
                 0 & 0 & -\mu ^2 & m_2^2 & 0 & 0 & 0 & 0 \\
                 0 & 0 & 0 & 0 & -\frac{\mu ^4}{m_2^2} & -\mu ^2 & 0 & 0 \\
                0 & 0 & 0 & 0 & -\mu ^2 & -m_2^2 & 0 & 0 \\
                 0 & 0 & 0 & 0 & 0 & 0 & \frac{3 \mu ^4}{m_2^2}-2 m_1^2 & \mu ^2 \\
                 0 & 0 & 0 & 0 & 0 & 0 & \mu ^2 & m_2^2 \\
            \end{array}
            \right) .
        \end{equation}
        
\begin{figure}[h]
        	\centering         
        	\begin{minipage}[b]{0.52\textwidth}           \includegraphics[width=\textwidth]{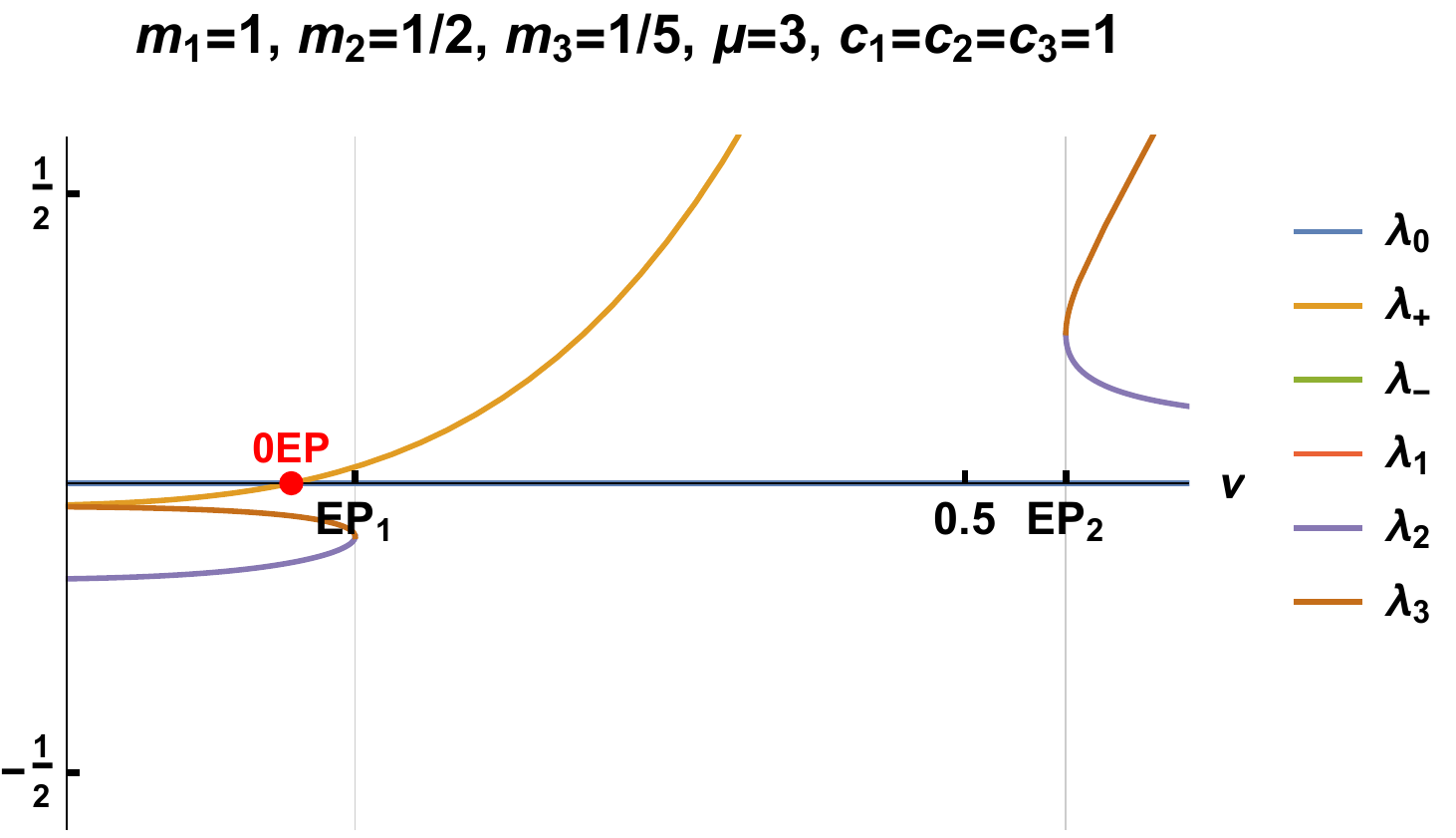}
        	\end{minipage}   
        	\begin{minipage}[b]{0.4\textwidth}           
        		\includegraphics[width=\textwidth]{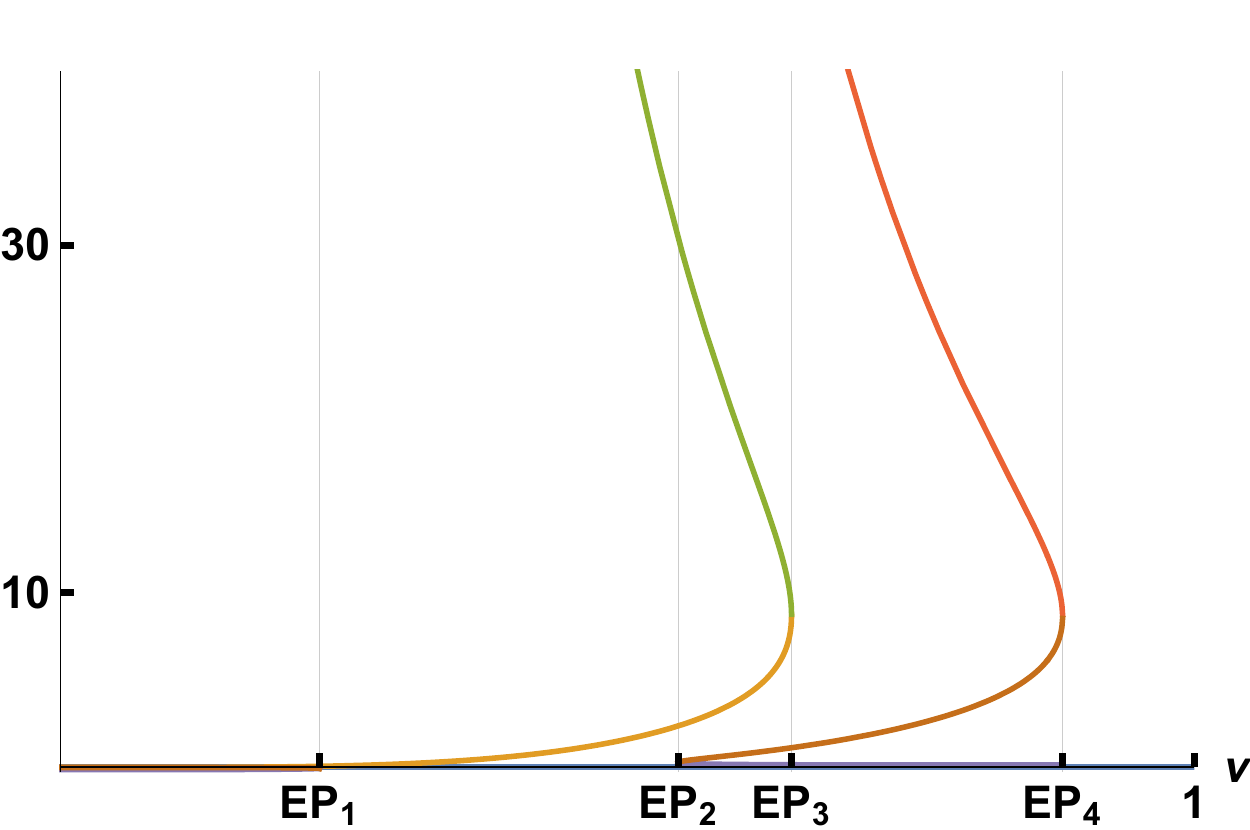}
        	\end{minipage}    
        	\caption{Real parts of the six eigenvalues $\lambda_0$, $\lambda_1$, $\lambda_2$, $\lambda_3$ and $\lambda_{\pm}$  of $M_{U(1),3}^{2}$ as functions of $\nu$, with all other parameters fixed as indicated, in the $U(1)$-broken regime. $\lambda_0=0$ corresponds to the Goldstone boson. Panel (b) is a zoomed out version of Panel (a).}
        	\label{fig:Eigenvalue spectrum of U(1) matrix}
        \end{figure}

        Key features of the model can be understood from the eigenvalue spectrum of $M_{U(1),3}^{2}$ plotted in figure \ref{fig:Eigenvalue spectrum of U(1) matrix} as a function of $\nu$ with all other parameters fixed as indicated. 
        Crucially we recognise the predicted number of Goldstone bosons to emerge, i.e. one corresponding to the eigenvalue $\lambda_0=0$.
        
        Moreover we observe in figure \ref{fig:Eigenvalue spectrum of U(1) matrix} panel (a) and (b) a standard feature in the eigenvalue spectra for complex matrices, namely that points in the parameter space at which two eigenvalues coincide and become complex conjugate pairs beyond this point. At these points also the eigenvectors coalesce and therefore the matrices are no longer diagonalisable. For energy spectra these points are commonly referred to as {{\em{exceptional points}} \cite{Kato1995}. However, unlike as for energy spectra, we have to exclude here the negative values in the $M^2$-spectrum to obtain physical, that is real, masses. Thus for instance the exceptional points $EP_1$ and $EP_2$ in figure \ref{fig:Eigenvalue spectrum of U(1) matrix} panel (a) are in the non-physical and physical regime, respectively.

        In addition we identify another special point in figure \ref{fig:Eigenvalue spectrum of U(1) matrix} at which the forbidden level crossing in Hermitian theories \cite{vonNeumann} is circumvented. Our analysis shows that at the point marked as 0EP the non-zero eigenvalue coalesces with the one for the zero eigenvalue. Moreover, just as for the standard exceptional point
        the two eigenvalues coalesce so that the matrix is non-diagonalisable. Thus in principle this point has all the trademarks of an exceptional point except for the fact that beyond this point, in both directions, the two eigenvalues do not become complex conjugate pairs but remain real as can be seen in figure \ref{fig:Eigenvalue spectrum of U(1) matrix} panel (a). Thus in order to distinguish the behaviour at these type of points from the one of standard exceptional points, we have termed this point as {\em{zero exceptional point}}. 
        
        Next we address the question of identifying whether the Lagrangians $\mathcal{L}_n^{SU(N)}$ are indeed physically meaningful. As a first exclusion principle we identify regions in the parameter space for which the model has a well-defined classical mass spectrum. Interestingly for $M_{SU(2),2}^{2}$ with the two choices $c_1=c_2=\pm 1$ no such region exists and these versions of the model must therefore be discarded as non-physical. However, for $c_1=-c_2=\pm 1$ we can identify physical regions in the parameter space as seen in figure \ref{fig: Physical reigion of SU(2) matrix} where these regions are plotted with respects to the parameter combinations $\mu^4 / m_1^4$ and $m_2^2 / m_1^2$. Of course there might be other arguments, also resulting at the quantum level, that force these models to be non-physical, but they pass the basic test of possessing a well defined classical mass spectrum.  
        
        \begin{figure}[h]
        	\centering
        	\begin{minipage}[b]{0.52\textwidth}           \includegraphics[width=\textwidth]{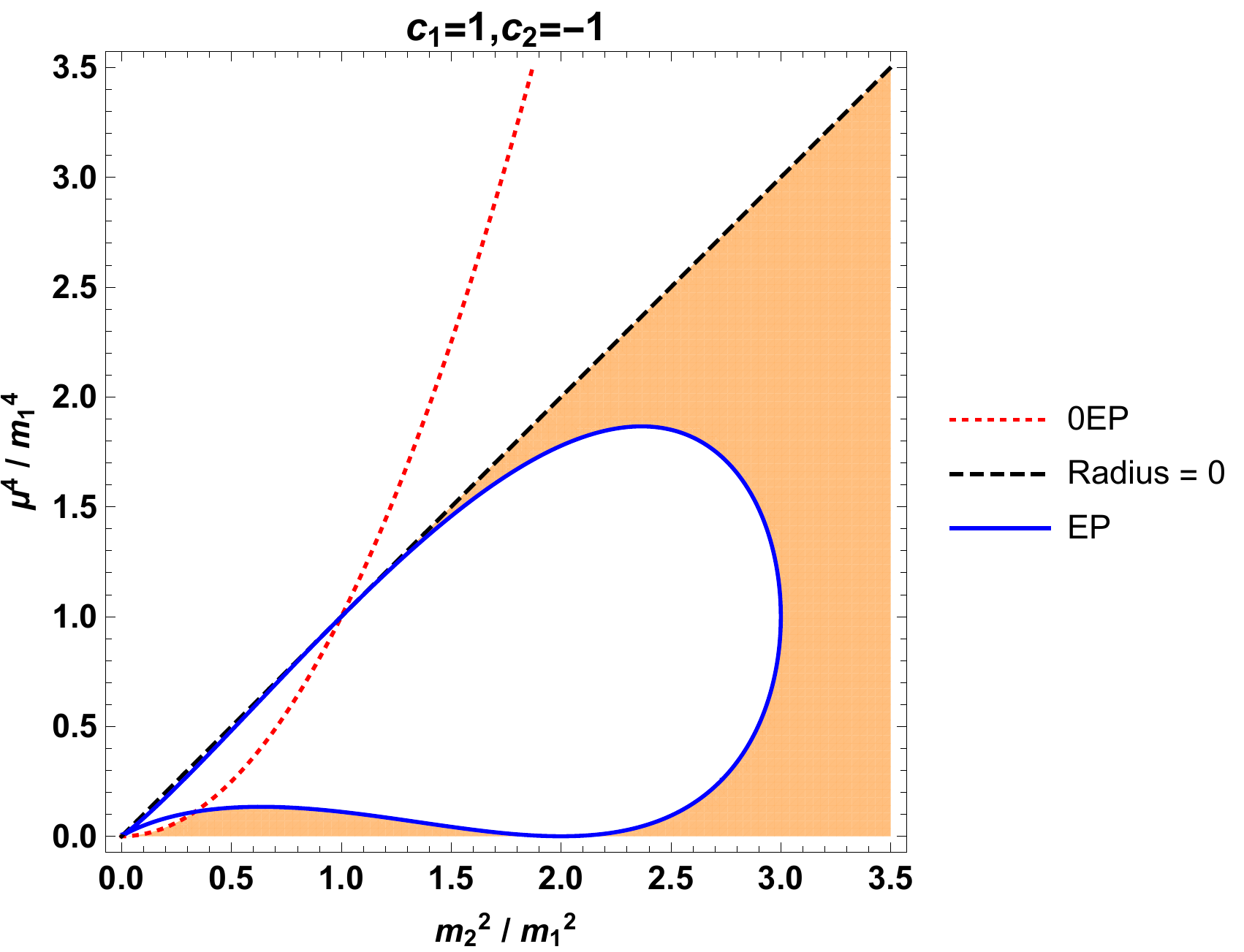}
        	\end{minipage}
        	\begin{minipage}[b]{0.4\textwidth}           \includegraphics[width=\textwidth]{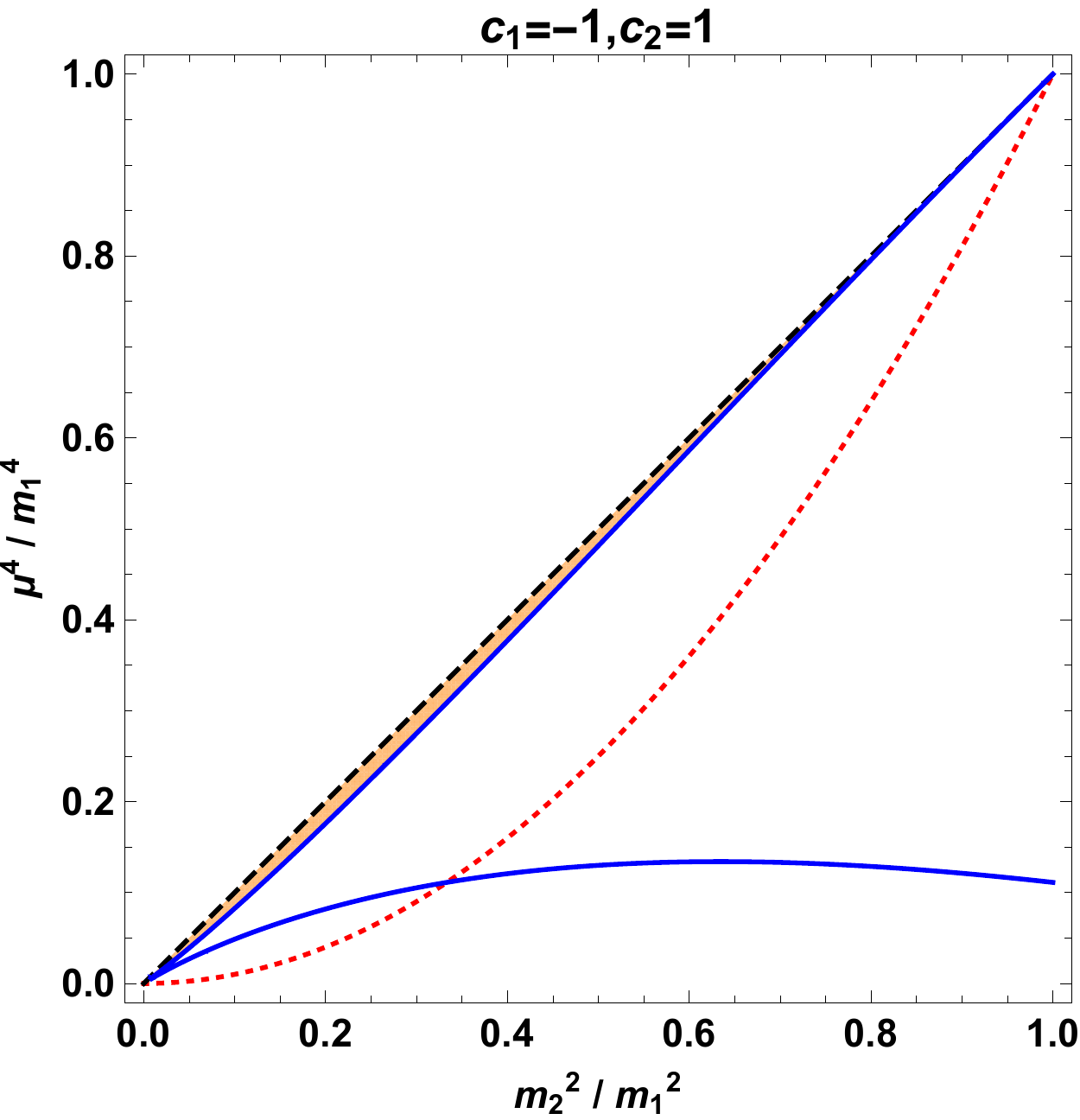}
        	\end{minipage}
        	\caption{Physical regions (in orange), resulting from $M_{SU(2),2}^{2}$ in the $SU(2)$-broken regime, at which the masses are positive definite bounded by exceptional points, zero exceptional points and trivial vacua with radius $R=0$ for different choices of the constants $c_i$.}
        	\label{fig: Physical reigion of SU(2) matrix}
        \end{figure}

        Having discussed the explicit forms of the physical regions and their boundaries, let us return to the question of whether the Goldstone bosons may be identified in the non-Hermitian theories considered. As shown in detail in \cite{FringGoldstone}, in the $\mathcal{CPT}$-symmetric regime the Goldstone theorem holds in the usual sense and the fields corresponding to the Goldstone bosons can be identified explicitly. Remarkably this is still possible at the standard exceptional point, despite the fact that mass squared matrix can not be diagonalised. Technically this needs to be done differently by exploiting the fact that the mass-squared matrix is in a block diagonal form, so that the block matrix with null eigenvector can still be diagonalisable at the exceptional point of other block matrices. Hence we can still identify the Goldstone fields at standard exceptional point. However, this is no longer possible at the zero exceptional point where the mechanism breaks down and the fields corresponding to the Goldstone boson can not be constructed. 
        
       Finally let us identify the $\mathcal{CPT}$-operators that ensure the reality of the mass-squared matrix and see whether they leave the complex Lagrangian $\mathcal{L}_n^{SU(N)}$ invariant. Using the definitions (\ref{definition of P}) and
       (\ref{definition of C}) for these operators, we solve the equations they must obey with the non-Hermitian Hamiltonian replaced by the non-Hermitian mass-squared matrix. As a concrete example we take the $3\times 3$ block matrix of the mass-squared matrix in the $n=3, U(1)$ model before the similarity transformation has been carried out, and solve all the property equations (\ref{properties of P}) and (\ref{three algebraic conditions for C}) for $\mathcal{P}$ and $\mathcal{C}$.
        Given the eigenvalues of the block matrix $\{\lambda_j\}=\{0,\lambda_- ,\lambda_+\}$, we    define the quantities $\Lambda_j := \lambda_j  +c_2 m_2^2 +c_3 m_3^2 $ and $\Lambda_j^k := \lambda_j +c_k m_k^2$. Then we obtain the solutions for $\mathcal{P}$ and $\mathcal{C}$ that can be written as
        \begin{equation}
            \mathcal{P}' = \sum_{j=0,\pm} \frac{s_j}{N_j^2} \left(\begin{array}{ccc}
                 \left(\Lambda_j^2 \Lambda_j^3 +\nu^4\right)^2& i\mu^2 \Lambda_j^3 \left(\Lambda_j^2 \Lambda_j^3 +\nu^4\right) &\mu^2 \nu^2\left(\Lambda_j^2 \Lambda_j^3 +\nu^4\right) \\
                 -i\mu^2 \Lambda_j^3 \left(\Lambda_j^2 \Lambda_j^3 +\nu^4\right)& \mu^4 \left(\Lambda_j^3\right)^2&-i\nu^2 \mu^4 \Lambda^3_j\\
                 \mu^2 \nu^2\left(\Lambda_j^2 \Lambda_j^3 +\nu^4\right) &i\nu^2 \mu^4 \Lambda^3_j& \mu^4 \nu^4
            \end{array}\right), \label{Pd}
        \end{equation}
        \begin{equation}
            \mathcal{C}' = \sum_{j=0,\pm} \frac{(-1)^{\delta_{-,j}}s_j}{N_j^2}  \left(\begin{array}{ccc}
                 \left(\Lambda_j^2 \Lambda_j^3 +\nu^4\right)^2& i\mu^2 \Lambda_j^3 \left(\Lambda_j^2 \Lambda_j^3 +\nu^4\right) &\mu^2 \nu^2\left(\Lambda_j^2 \Lambda_j^3 +\nu^4\right) \\
                 i\mu^2 \Lambda_j^3 \left(\Lambda_j^2 \Lambda_j^3 +\nu^4\right)& -\mu^4 \left(\Lambda_j^3\right)^2&i\nu^2 \mu^4 \Lambda^3_j\\
                 \mu^2 \nu^2\left(\Lambda_j^2 \Lambda_j^3 +\nu^4\right) &i\nu^2 \mu^4 \Lambda^3_j& \mu^4 \nu^4
            \end{array}\right). \label{Cd}
        \end{equation}
        The normalisation constants are $N_0^2=\kappa \lambda_- \lambda_+$, $N_\pm^2=(\kappa +\lambda_\pm \Lambda_\pm) \lambda_\pm (\lambda_+-\lambda_-)$. As the solutions depend on the choices for the set of signatures $s_0,s_\pm = \pm 1$, we have found eight different possible solutions for these operators. The two particular parity operators  $\mathcal{P} (s_0=\pm,s_-=\mp,s_+=\pm)$ are rather special because they can be identified as the operators involved in the $\mathcal{CPT}_{1/2}$ transformations in (\ref{symm}) that leave also the entire Lagrangian invariant when switching on the interaction terms. We still have an ambiguity of three operators left, which reflects the aforementioned non-uniqueness of the metric and the Dyson map in the $\mathcal{C}'$ and $\mathcal{P}'$-operators.

    \subsection{Higgs mechanism in a model with local $SU(2)$ symmetry}
        While in the previous section the continuous symmetry has been global, in the sense of being the same at all points in spacetime, we will now allow these to depend on points in spacetime, that is to be local. Via the standard mechanism of minimal coupling this will introduce local gauge fields and in turn is also expected to remove the degrees of freedom of the Goldstone bosons. The latter phenomena is known as the celebrated Higgs mechanism \cite{Higgs_mechanism_ref_1,Higgs_mechanism_ref_2,Higgs_mechanism_ref_3,Higgs_mechanism_ref_4} in the standard Hermitian setting. In \cite{FringHiggs} we showed that the mechanism is also extendable to non-Hermitian theories, although differently in different regimes characterized by the $\mathcal{CPT}$-symmetry as we shall discuss. We illustrate this for a model with local $SU(2)$ symmetry of the form
        \begin{equation}\label{local su(2) action}
            \mathcal{L}_l=\sum_{i=1}^2 \left|D_\mu \phi_i\right|^2 +m_i^2 \left|\phi_i\right|^2 -\mu^2 \left(\phi_1^\dagger \phi_2 -\phi_2^\dagger \phi_1 \right)-\frac{g}{4} \left(\left|\phi_1\right|^2\right)^2 -\frac{1}{4}F_{\mu\nu}F^{\mu\nu}.
        \end{equation}
        This Lagrangian is identical to $\mathcal{L}_{SU(2)}^{n=2}$ introduced in the previous section, but minimally coupled with local gauge fields $A_\mu$, so that partial derivatives are replaced by covariant derivatives $D_\mu = \partial_\mu - i e A_\mu$ with charge $e \in \Bbb{R}$. A standard Yang-Mills term comprised of the Lie algebra
        valued field strength $F_{\mu\nu}= \partial_\mu A_\nu - \partial_\nu A_\mu -i e [A_\mu,A_\nu]$ has also been added. The two anti-linear symmetries in (\ref{symm}) are still respected by $\mathcal{L}_l$ as in the previous section. 
        
        We may also still map $\mathcal{L}_l$ to an equivalent Hermitian Lagrangian by means of an adjoint action of the Dyson map (\ref{DM}), albeit with changed conjugate momenta due to the presence of the gauge field. With this modification the field transformations due to the action of the Dyson map are unchanged. The potential term of the resulting real Lagrangian is equivalent to the one in $\mathcal{L}_{SU(2)}^{n=2}$, so when expanding the potential around the non-trivial symmetry breaking vacuum we find same mass-squared matrix as given in (\ref{Global SU(2) mass matrix}).
        
        Expanding also the kinetic term around the non-trivial symmetry breaking vacuum, we \cite{FringHiggs} found the following mass for the gauge vector boson
        \begin{equation}\label{Gauge mass}
            m_g = \frac{e R_f}{m_2^2} \sqrt{m_2^4 -\mu^4} ,
        \end{equation}
        where $R_f=\sqrt{4(\mu^4 + c_1 c_2 m_1^2 m_2^2)/gm_2^2}$ is the radius of the spherical vacuum solution in the fundamental representation.
        This gauge mass is real and positive when $R_f$ and $\sqrt{m_2^4 - \mu^4}$ are simultaneously positive definite. Thus in these cases the Higgs mechanism works in the same fashion as in the Hermitian case and the gauge vector boson becomes massive. However, there are now several interesting scenarios where this might be violated. First we note that the vacuum radius can never be negative for  $c_1 = c_2 = \pm 1$, but these two cases have been excluded previously as being non-physical. Therefore the vacuum radius always takes the form $R_f=4(\mu^4 -m_1^2 m_2^2)/gm_2^2$ for the two remaining physical cases. Next we notice that the scenario with vanishing $R_f$ for $\mu^4 =m_1^2 m_2^2$ corresponds to the black dotted line in both panels of figure \ref{fig: Physical reigion of SU(2) matrix}. At these lines the theory is expanded around the trivial symmetric vacuum so that no Goldstone bosons emerge that may be combined with the gauge field. The square root factor in (\ref{Gauge mass}) is more interesting as it may be tuned to give rise to novel phenomena only present in non-Hermitian theories. When $m_2^4 =\mu^4$ the mass of the gauge field also vanishes. This scenario corresponds to the red dotted line in both panels in figure \ref{fig: Physical reigion of SU(2) matrix}, that is to the zero exceptional points. As we discussed in the previous section, at these points the Goldstone boson field can not be identified even though the zero mass eigenvalue is correctly predicted by Goldstone's theorem and found in the eigenvalue spectrum of the mass-squared matrix. This means there is no well-defined Goldstone boson that can be combined with the gauge field to ensure that it acquires a non-vanishing mass. Thus {\em{the Higgs mechanism breaks down at the zero exceptional point with the Goldstone boson being unidentifiable and the gauge particle unable to acquire a mass.}} Interestingly, with regard to the Higgs mechanism  nothing special happens at the exceptional points which constitute the remaining boundary of the physical region indicated in figure \ref{fig: Physical reigion of SU(2) matrix} by solid blue lines. Despite the fact that the Goldstone boson field has to be constructed technically in a different fashion, it can actually be identified and hence can be combined with the gauge field to give the latter a non-vanishing mass.       
        
        Having discussed the breakdown of the Higgs mechanism at two types of boundaries of the physical regions as indicated by the orange region in figure \ref{fig: Physical reigion of SU(2) matrix}, let us now briefly explain how the Higgs mechanism works inside the physical region of our non-Hermitian model.
 Rewriting for this purpose the Lagrangian in terms of the Goldstone fields $\{G^a\}$, we found in \cite{FringHiggs} the following expression for the kinetic term
        \begin{eqnarray}
            \mathcal{L} &=& \sum_{a=1}^3 \partial_\mu G^a \partial^\mu G^a - m_g A_\mu^1 \partial^\mu G^1 + m_g A_\mu^2 \partial^\mu G^1+m_g A_\mu^3 \partial^\mu G^3 + \frac{1}{2}m_g^2 A_\mu^a A^{a\mu}+\dots  \nonumber\\
            &=& \frac{1}{2}m_g^2 \left(A_\mu^1 - \frac{1}{m_g} \partial_\mu G^1\right)^2 +\frac{1}{2}m_g^2 \left(A_\mu^2 + \frac{1}{m_g} \partial_\mu G^2\right)^2+\frac{1}{2}m_g^2 \left(A_\mu^3 + \frac{1}{m_g} \partial_\mu G^3\right)^2+\dots \nonumber\\
            &=& \frac{1}{2} m_g^2 \sum_{a=1}^3 B_\mu^a B^{a\mu}+\dots.
        \end{eqnarray}
        Thus by introducing the new gauge field $B_\mu^a=A_\mu^a \pm \frac{1}{m_g} \partial_\mu G^a$ we have removed the degrees of freedom of the Goldstone bosons and introduced a mass term for these combined gauge field-Goldstone boson fields.
        Evidently when the Goldstone boson fields can not be found, as at the zero exceptional points, or do not exist at all, as in a theory expanded around a trivial symmetry preserving vacuum, and the new $B_\mu$ fields can not be defined. Remarkably at the standard exceptional points this is still possible, when adapting the mechanism to identify the Goldstone bosons appropriately as explained in the previous section.   
        
        We note that the analysis presented in this section can be extended to the $SU(N)$-symmetric model in a straightforward fashion for which we found $N^2 -1$ gauge fields with the same masses as given in (\ref{Gauge mass}) and the same behaviour in all regions of the parameter space.
\section{Complex BPS solitons and magnetic monopoles with real energy}\label{Section: complex BPS solutions with real energy}

Non-Abelian gauge theories are known to possess an intriguing variety of different types of solutions to their equations of motion, such as solitons and almost unavoidably also magnetic monopole solutions \cite{tHooft_monopole,Polyakov_monopole,goddard1978magnetic}. More than fourty years ago Olive and Montonen \cite{montonen1977magnetic} noticed that different types of solutions may be related to each other and that the soliton solutions in non-Abelian gauge theories become equivalent to massive gauge fields in a dual theory.
Here we wish to address the question of whether these type of features will survive, or in which way they need to be altered, in a non-Hermitian theory \cite{FringMonopole} and pay special attention to whether the energies of the solutions might still be real.

Extremely instrumental in finding explicit analytic solutions is the  Bogomol'nyi-Prasad-Sommerfield (BPS) \cite{Bogomolny_scaling_limit,Prasad_Sommerfield_scaling_limit}
limit, consisting of taking a multi-scaling limit in a physically motivated fashion by acknowledging that certain mass ratios are very small. While most of the equations in this context are very complicated coupled differential equations, this approach allows to convert to a solvable system with explicit solutions that, however, still contains the key features of the physics involved. In the next subsection we will first demonstrate how these limits may be taken to obtain t'Hooft-Polyakov monopole solutions and subsequently consider a general setting of theories for which the limit is assumed to have been already carried out in 1+1 and also 3+1 dimensions.  

For our solutions to be physically meaningful we require a reality condition on them as well as on the Hamiltonian so that the corresponding energies become real. A simple argument that adapts the reality condition from the quantum mechanical to the field theory \cite{Fring_reality_condition_ref} is simply demanding the Hamiltonian and field solution to be $\mathcal{PT}$-symmetric. When
\begin{itemize}
	\item[(a)] the Hamiltonian satisfies $\mathcal{H}[\phi(x_\mu)]=\mathcal{H}^\dagger[\phi(-x_\mu)]$, and
	\item[(b)] the solutions to the equations of motion are $\mathcal{PT}$ symmetric, $\mathcal{PT} \phi(t,\Vec{x})=\phi(-t,-\Vec{x})=\phi(t,\Vec{x})$,
\end{itemize}
then the energy $E$ of the solutions $\phi$ is real. This follows from the simple argument
\begin{equation}
	E= \int^\infty_{-\infty}dx \mathcal{H}[\phi(x)]=-\int_\infty^{-\infty}dx \mathcal{H}[\phi(-x)]=\int^\infty_{-\infty}dx \mathcal{H}^\dagger [\phi(x)]=E^* .
\end{equation}
In \cite{FringBPS,FringSkyrmion} we found that these requirements need to be extended to include more possibilities when there are non-trivial anti-linear symmetry relating two degenerate solutions. In such a scenario we replace the $\mathcal{PT}$ symmetry of the solution with a generic anti-linear symmetry that we refer to as modified $\mathcal{CPT}$, which is not to be confused with the standard $\mathcal{CPT}$ in quantum field theory. Then the reality of the energy is guaranteed if the following three conditions hold:
\begin{itemize}
	\item[\textbf{(i)}] The Hamiltonian transforms under the modified $\mathcal{CPT}$-symmetry as
	\begin{equation}\label{condition i}
		\mathcal{CPT}:\mathcal{H}[\phi (x_\mu)]\rightarrow \mathcal{H}^\dagger[\phi (-x_\mu)].
	\end{equation}
	\item[\textbf{(ii)}] Two solutions to the equations of motion $\phi_1$ and $\phi_2$, not necessarily distinct, are related to each other by the $\mathcal{CPT}$-symmetry as
	\begin{equation}\label{condition ii}
		\mathcal{CPT}: \phi_1 (x_\mu) \rightarrow  \phi_2 (-x_\mu).
	\end{equation}
	The $\mathcal{CPT}$ symmetry is here a generic anti-linear symmetry which may or may not coincide with the $\mathcal{PT}$ symmetry.
	\item[\textbf{(iii)}] The energies $E[\phi]$ of the two solutions are degenerate
	\begin{equation}\label{condition iii}
		E[\phi_1]=E[\phi_2].
	\end{equation}
\end{itemize}
When $\phi_1 = \phi_2$ and $\mathcal{CPT}=\mathcal{PT}$, the condition (i), (ii) and (iii) coincide with (a) and (b). While for a quantum mechanical systems the anti-linear symmetries can usually be read off trivially from the Hamiltonian and are then simply verified for the wave functions in the $\mathcal{PT}$-symmetric regime, here the starting point is mostly reversed and it appears more practical to identify the relevant symmetry from the explicit solutions first. 

For N-soliton solutions these reality conditions have to be enlarged by employing also the integrability of the model and using the fact that asymptotically the N-soliton solutions separate into N one-soliton solutions \cite{CenFring,cen2016time}.

Let us now construct concrete solutions in some explicit non-Hermitian systems and study their properties.

    \subsection{t'Hooft-Polyakov magnetic monopoles}
        In order to study how complex monopole solutions might arise we consider a complex extended version of a Lagrangian \cite{FringMonopole} for which real t'Hooft-Polyakov monopole solutions were shown to exist \cite{tHooft_monopole,Polyakov_monopole} by adding a complex part in a similar fashion as in our model used to investigate the non-Hermitian version of the Higgs mechanism in (\ref{local su(2) action})
        \begin{eqnarray}\label{Lagrangian of monopole}
            \mathcal{L}_{cm} &=& \frac{1}{2} Tr\left(D \phi_1 \right)^2+\frac{1}{2} Tr\left(D \phi_2\right)^2-c_1 m_1^2 Tr\left( \phi_1^2\right) +c_2 m_2^2 Tr\left( \phi_2^2\right)\\
            && -i\mu^2 Tr\left(\phi_1 \phi_2\right) -\frac{g}{4} Tr \left(\phi_1^2\right)^2 -\frac{1}{4}Tr\left(F_{\mu\nu}F^{\mu\nu}\right) . \nonumber
        \end{eqnarray}
        Here we take $\phi_1$ and $\phi_2$ to belong to the adjoint representation of $SU(2)$, which means we can factorise the fields as $\phi_i (x) = \phi_i^a (x) \tau^a$, where $\phi_i^a (x)$ are a real fields and $\{\tau^a \}$ denote the generators of $SU(2)$. Noting that the adjoint action of the Dyson map 
        \begin{equation}
            \eta_\pm =\prod_{a=1}^3 \exp\left(\pm \frac{\pi}{2}\int d^3 \Pi^a \phi_2^a\right)
        \end{equation}
        transforms the fields as $\phi_1 \rightarrow \phi_1$, $\phi_2 \rightarrow c_3 i \phi_2$ with $c_3 = \pm 1$, we can map the complex monopole Lagrangian $\mathcal{L}_{cm}$ to two equivalent Hermitian ones. The equations of motion and the Higgs vacuum may then be found in a standard fashion as described above. Next we assume the parametrization
        \begin{equation}
        	(\phi _{\alpha }^{cl})^{a}=h_{\alpha }(r)\hat{r}_{n_{\alpha
        	}}^{a}~,~~~(A_{i}^{cl})^{a}=\epsilon ^{iaj}\hat{r}_{n}^{j}\left( \frac{u(r)-1%
        	}{er}\right),~~~ \hat{r}_n^a =\!\!   \left(%
        \!\! {\footnotesize \begin{array}{c}
        	\sin(\theta)\cos(n \varphi) \\ 
        	\sin(\theta)\sin(n \varphi) \\ 
        	\cos(\theta)%
    \end{array}} \!\!
\right)   \label{par1}
        \end{equation}
    for the scalar and the gauge fields, respectively, with $n$ denoting the winding number. This Ansatz is motivated by the fact that the static solutions have to converge to the vacuum solution in order to have finite energy \cite{goddard1978magnetic,derrick1964comments}
    \begin{equation}\label{boundary condition of Monopole}
    	\lim_{r\rightarrow \infty} h_1(r) = \pm R_a ~,~~~\lim_{r\rightarrow \infty} h_2(r) = \mp \frac{c_2 c_3 \mu^2}{m_2^2} R_a,
    \end{equation}
    with $R_a=\sqrt{(m_1^2 m_2^2 - \mu^4)/2g m_2^2}$ denoting the radius of the spherical vacuum solution in the adjoint representation. The equations of motion then acquire the form
        \begin{eqnarray}
        	u^{^{\prime \prime }}(r)+\frac{u(r)\left[ 1-u^{2}(r)\right] }{r^{2}}+\frac{%
        		e^{2}u(r)}{2}\left\{ h_{2}^{2}(r)-h_{1}^{2}(r)\right\}  &=&0,
        	\label{mono1} \\
        	h_{1}^{^{\prime \prime }}(r)+\frac{2h_{1}^{^{\prime }}(r)}{r}-\frac{%
        		2h_{1}(r)u^{2}(r)}{r^{2}}+g\left\{ c_{1}\frac{m_{1}^{2}}{g}h_{1}(r)+c_{3}%
        	\frac{\mu ^{2}}{g}h_{2}(r)+2h_{1}^{3}(r)\right\}  &=&0,
        	\label{mono2} \\
        	h_{2}^{^{\prime \prime }}(r)+\frac{2h_{2}^{^{\prime }}(r)}{r}-\frac{%
        		2h_{2}(r)u^{2}(r)}{r^{2}}+c_{2}m_{2}^{2}\left\{ h_{2}(r)+c_{3}\frac{\mu ^{2}%
        	}{m_{2}^{2}}h_{1}(r)\right\}  &=&0,
        	\label{mono3}
        \end{eqnarray}
       extending the standard set of equations previously obtained in \cite{tHooft_monopole,Polyakov_monopole} by a new field $h_{2}$ and the additional equation (\ref{mono3}). Evidently these equations are difficult to solve. In order to facilitate the construction of explicit analytic solutions we derive the BPS equations by carrying out the fourfold limit   
        \begin{equation}
            \lim\limits_{g,m_1,m_2,\mu \rightarrow 0} \!\!\!\!\!\! \text{(\ref{mono1}),(\ref{mono2}),(\ref{mono3})} \quad \text{with} \quad X:=\frac{m_1^2}{g}<\infty ,  Y:=\frac{\mu^2}{g}<\infty, Z:=\frac{\mu^2}{m_2^2}<\infty.
        \end{equation}
     Remarkably the equations obtained in this manner may be solved explicitly. Adjusting the boundary conditions appropriately, as indicated in (\ref{boundary condition of Monopole}), we obtain two distinct sets of solutions, which connect two different vacua
    \begin{eqnarray} \label{Explicit solitions of monopole}
    		u(r) &=& \pm \frac{erlR_a}{ \sinh (erlR_a)},\\
    	h_1^\pm (r)&=& \pm \text{Sign} (n) \frac{ 1}{l}\left\{|lR_a|\coth\left(e|lR_a|r\right) -\frac{1}{er}\right\},\\
    	h_2^\pm &=& \mp \text{Sign} (n) \frac{ c_2 c_3 Z}{l}\left\{|lR_a|\coth\left(e|lR_a|r\right) -\frac{1}{er}\right\}, \label{ex3}
    \end{eqnarray}
    with $l := \sqrt{1 -Z^2}$ and $R_a=\sqrt{(X-YZ)/2}$.   
    
    Let us now consider whether the energies for these solutions are real despite the fact that the Hamiltonian of the system is non-Hermitian. After the similarity transformation has been carried out, the energy for a static solution to our system $\mathcal{L}_{cm}$  is given by 
    \begin{equation}
    	E = \int d^3 x Tr \left(B^2\right) +Tr\left\{(D_i \phi_1)^2\right\}-Tr\left\{(D_i \phi_2)^2\right\}+V , \label{En}
    \end{equation}
    which when evaluated for the two solutions in (\ref{Explicit solitions of monopole})-(\ref{ex3}) leads to the same energies for both 
        \begin{equation} 
            E=\frac{8|n|\pi R_a}{e} \left(\frac{1 - Z^2}{\sqrt{1 - Z^2}}\right)=\frac{8|n|\pi R_a}{e}l. \label{MonE}
        \end{equation}
       
       Next we compare the behaviour of the gauge mass, given in (\ref{Gauge mass}) with $R_f \rightarrow R_a$ as derived in \cite{FringMonopole}, versus the monopole mass taken to be the rest mass, i.e.~it is equal to its energy E in (\ref{MonE}). In figure \ref{fig: gauge vs monopole masses 2} we notice that in the weak coupling regime for $e=2$ we have $m_g < m_m$, with exchanged ordering in the strong regime for $e=10$. Most notably, however, is the behaviour seen in panels (a) and (b) showing that both masses vanish simultaneously at the zero exceptional point, where they both become purely imaginary and hence non-physical. As functions of $Z$ they revive as real valued beyond the trivial vacuum for which $R_a=0$. In panels (c) and (d) we show that the non-physical region can be made to vanish when the zero-exceptional point and the trivial vacuum coincide, which is the case for the choice $Z = 1$.

        \begin{figure}[h]
            \centering
            \begin{minipage}[b]{1\textwidth}      \includegraphics[width=\textwidth]{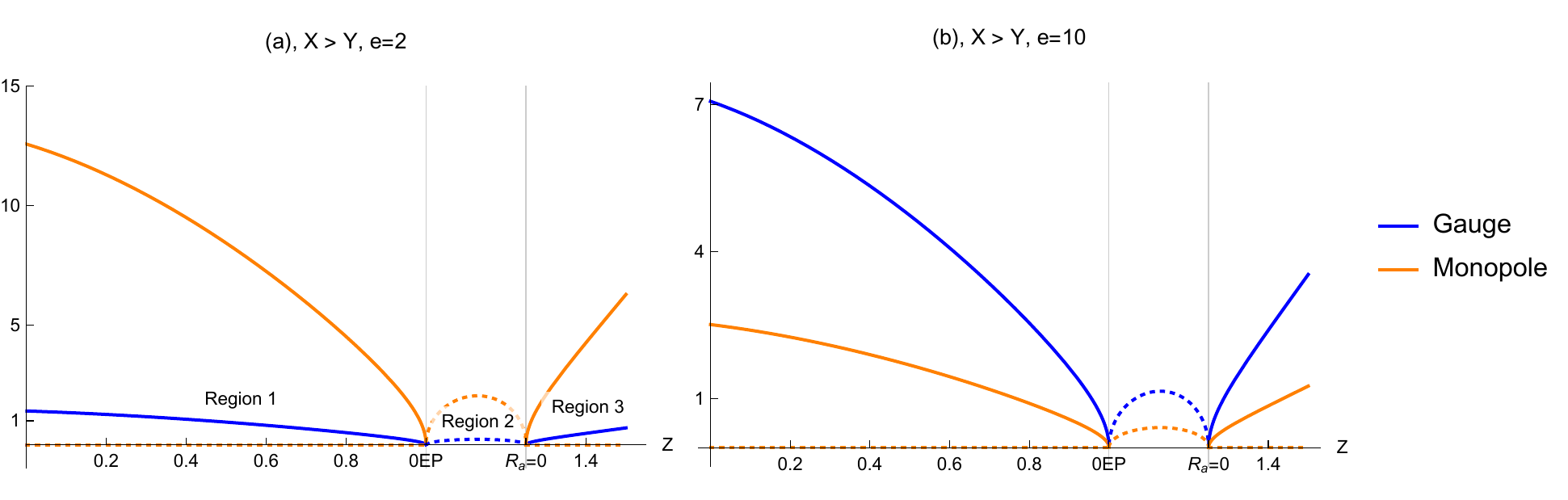}
            \end{minipage}
            \begin{minipage}[b]{1\textwidth}      \includegraphics[width=\textwidth]{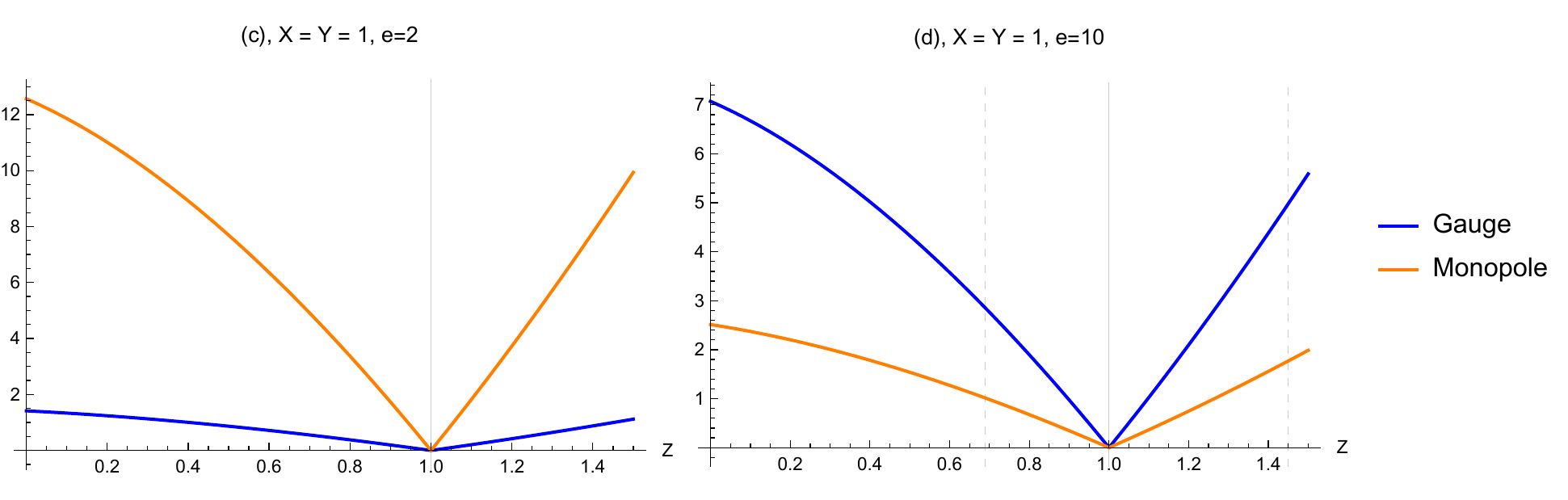}
            \end{minipage}
            \caption{Revival of the gauge and magnetic monopole masses in the weak and strong regime, panels (a), (c) and (b), (d), respectively. In panel (a) and (b) we have $X=1$, $Y=4/5$ and $n=1$. Solid lines correspond to real and dotted lines to imaginary parts.}
            \label{fig: gauge vs monopole masses 2}
        \end{figure}
        
        We conclude this section by explaining the reality of the energies in different regions employing the conditions (i)-(iii) in (\ref{condition i})-(\ref{condition iii}). First, we identify the $\mathcal{CPT}$-symmetry from the Lagrangian by noticing that the Hamiltonian resulting from $\mathcal{L}_{cm}$ is mapped to it conjugate under the transformation
 \begin{equation}\label{CPTsy}
 	\mathcal{CPT}:  \phi_i (x) \rightarrow (-1)^{\delta_{i2}} \left[\phi_i (-x)\right]^\dagger .
 \end{equation}          
        This constitutes the condition (i) in (\ref{condition i}). Acting with this symmetry on the two scalar fields in (\ref{par1}) we find condition (ii) realised as     
     \begin{eqnarray*}
     	{\cal CPT} :& &\phi _{1}^{\pm }(x)\rightarrow \left[ \phi _{1}^{\pm }(-x)%
     	\right] ^{\dagger }=\left\{ 
     	\begin{array}{ll}
     		\phi _{1}^{\mp }(x) & ~\text{in region~1} \\ 
     		\phi _{1}^{\pm }(x) & ~\text{in region~3}%
     	\end{array}%
     	 \right. ,~~ \\
     	&& \phi _{2}^{\pm }(x) \rightarrow -\left[ \phi _{2}^{\pm }(-x)\right]
     	^{\dagger }=\left\{ 
     	\begin{array}{ll}
     		\phi _{2}^{\pm }(x) & ~\text{in region~1} \\ 
     		\phi _{2}^{\mp }(x) & ~\text{in region~3}%
     	\end{array}%
     	\right. .
     \end{eqnarray*}%
     We used here the properties $\left( h_{1}^{\pm }\right) ^{\dagger
     }=h_{1}^{\pm }$, $\left( h_{2}^{\pm }\right) ^{\dagger }=h_{2}^{\mp }$ in
     region 1 whereas in region 3 we have $\left( h_{1}^{\pm }\right) ^{\dagger }=h_{1}^{\mp }$, $\left(
     h_{2}^{\pm }\right) ^{\dagger }=h_{2}^{\pm }$. In region 2 the conjugate fields $%
     \left( h_{1}^{\pm }\right) ^{\dagger }$ and $\left( h_{2}^{\pm }\right)
     ^{\dagger }$ are not solving the BPS equations. Hence in region 2 the ${\cal CPT}$-symmetry is broken and we therefore expect complex energies. In regions 1 and 3 we have always one of the fields mapped to itself, whereas the mapping of the other field exchanges dual and self-dual fields. Since
     the energies of the dual and self-dual fields are the same, also condition
     (iii) is satisfied so that the energies of the solutions must be real.

     \subsection{Dual BPS theories in 1+1 dimensions}  
     Motivated by the success of the theories obtained in the BPS limit, as exemplified in the previous subsection, one may take a different standpoint and regard the BPS theories as the starting point of one's considerations in their own right. For this purpose one may formulate a generic theory that possesses the general properties observed in BPS theories based on the original procedure \cite{Bogomolny_scaling_limit} by completing the square or other methods such employing the concept of strong necessary conditions \cite{st}. In \cite{Adam_BPS_procedure} the authors build on these observations by pointing out that the energy functional and topological charges of a BPS theory can be cast into the generic forms
        \begin{eqnarray}\label{General form of the BPS energy}
        	E= \int d^d x \left( A^2 + \Tilde{A}^2 \right)= \int d^d x \left[ (A\mp \Tilde{A})^2\pm2A\Tilde{A} \right] ~,~~~ Q=\int d^d x A\Tilde{A},
        \end{eqnarray}
    where the quantities $A$ and $\Tilde{A}$ are functions of the fields and their derivatives appearing in the defining Lagrangian of the theory. The energy in (\ref{En}) can for instance be shown to be of that precise form too when the BPS limit is carried out. Using the definitions in (\ref{General form of the BPS energy}) one may then show \cite{Adam_BPS_procedure} that the compatibility condition between the Euler-Lagrange equation resulting from functionally varying the energy $E$ and the topological charge conservation $\delta Q=0$ when varied with respect to a field change, leads to the anti/self-dual equations $A=\pm \Tilde{A}$. The latter equation is interpreted as the BPS equation, since it achieves at the same time that the charge $Q$ saturates the Bogomolny bound $E = |Q|$, as seen from the relations in (\ref{General form of the BPS energy}).
    
    Specifying the expression for $E$ and $Q$ further by introducing scalar fields $\phi$, a non-trivial target space metric $\eta$ and a pre-potential $U$, one may take them to be of the still fairly general form \cite{Ferreira_BPS_non-trivial_metric}  
        \begin{eqnarray}
        	E &=& \frac{1}{2}\int \partial_\mu \phi_a \eta_{ab}\partial^\mu \phi_b + \eta^{-1}_{ab}\frac{\delta U}{\delta \phi_a} \frac{\delta U}{\delta \phi_b} , \label{E1} \\
        	Q &=& \int_{\infty}^{\infty} dx \frac{\partial U }{\partial x} =\lim_{x\rightarrow \infty}U[\phi(x)] -\lim_{x\rightarrow -\infty}U[\phi(x)]  . \label{topological charge of prepotential}
        \end{eqnarray}
     Comparing with the solutions in (\ref{General form of the BPS energy}) one identifies
        \begin{equation}
        	A = \rho_{ab}\partial_x \phi_b ~,~~~ \Tilde{A} = \frac{\delta U}{\delta \phi_b}\rho_{ba}^{-1},
        \end{equation}
    where the metric is factorised as $\eta= \rho^T \rho$. Several examples for this setting that lead to real solutions were considered in \cite{Ferreira_BPS_non-trivial_metric}. Here we recall one of many examples of complex extensions studied in \cite{FringBPS} that led to complex solutions with real energies. 
        
        We make a complex choice for the target space metric and keep the pre-potential real
        
        \begin{equation}
            \eta=\left(\begin{array}{cc}
                 1&-i\lambda  \\
                 -i\lambda&1 
            \end{array}\right),  \qquad \text{and} \qquad U=-(\cos\phi_1 +\mu\phi_1+\cos\phi_2),
        \end{equation}
        with $\lambda, \mu \in \mathbb{R}$ and $\phi_1 (t,x), \phi_2 (t,x)$ being real scalar fields. The potential, being the last term in (\ref{E1}), resulting from these choices is a complex coupled sine-Gordon potential
        \begin{equation}\label{potential in 1+1 example}
            V= \frac{1}{2(1+\lambda^2)} \left[(\sin\phi_1 -\mu)^2 + 2i \lambda (\sin\phi_1 -\mu)\sin\phi_2 +\sin^2 \phi_2\right].
        \end{equation}
         In the Hermitian limit $\lambda=0$ we can solve the corresponding BPS equations analytically to 
        \begin{eqnarray}
            \phi_1^{\pm (n)} &=&2\text{arctan}\left\{\frac{1}{\mu}+\frac{\sqrt{1-\mu^2}}{\mu}\tanh\left[\frac{1}{2}\sqrt{1-\mu^2}(\pm x+\kappa_1)\right]\right\}+2n\pi\label{solution 1 of coupled sine-Gordon model},\\
            \phi_2^{\pm (n)} &=&2\text{arctan} \left(e^{\pm x + \kappa_2}\right)+2n\pi . \label{solution 2 of coupled sine-Gordon model}
        \end{eqnarray}
    
     \noindent In the non-Hermitian case, $\lambda \not=0$, we have to resort to solving the equations numerically. A set of sample kink and anti-kink solutions are depicted in figure \ref{fig: Numerical plot of kink solutions}. Crucially we observe that the asymptotic limits for the real parts do not depend on $\lambda$ and those for the imaginary parts are always vanishing. 
     
      \begin{figure}[h]
     	\centering
     	\begin{minipage}[b]{0.52\textwidth}    \includegraphics[width=\textwidth]{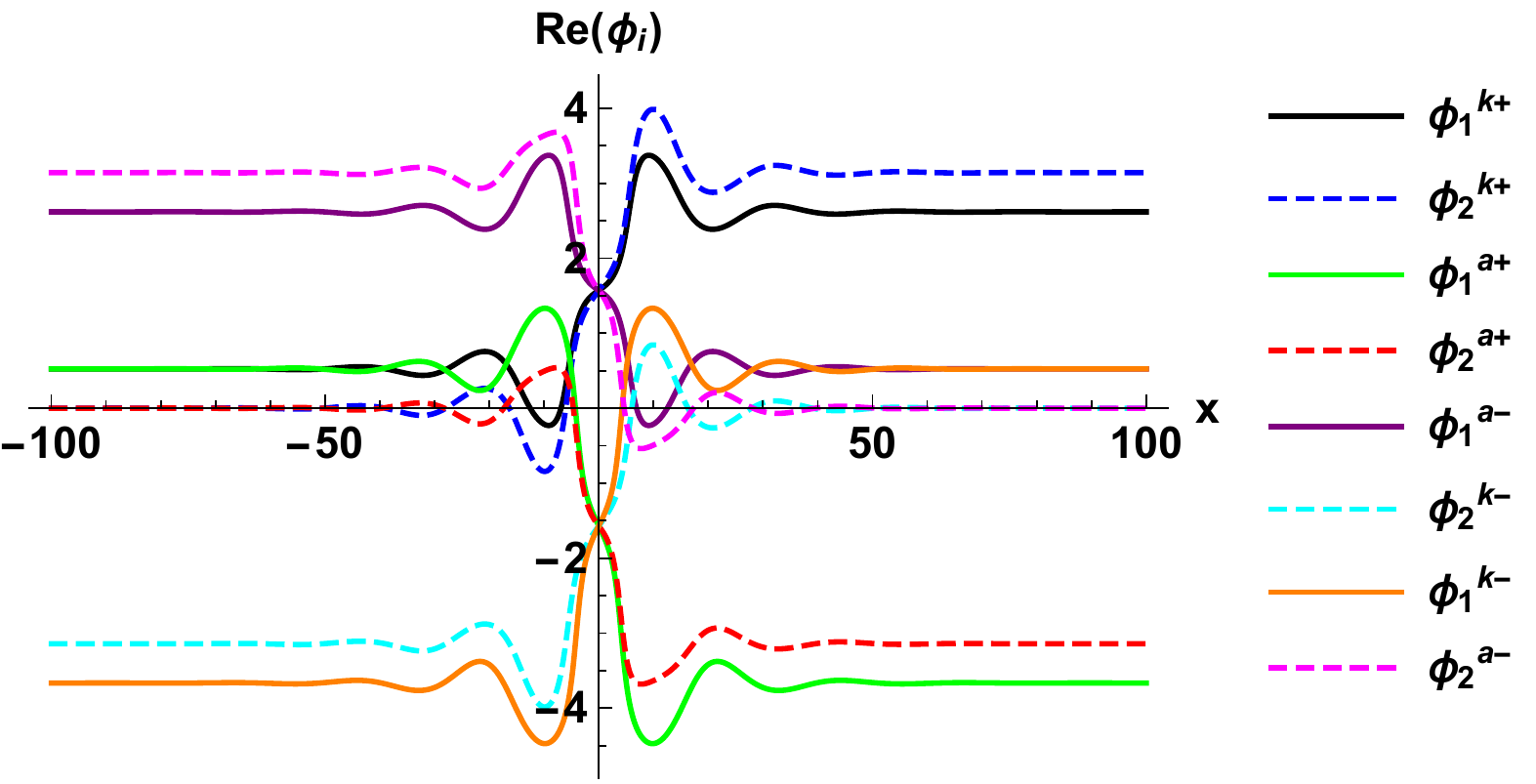}
     	\end{minipage}
     	\begin{minipage}[b]{0.4\textwidth}     \includegraphics[width=\textwidth]{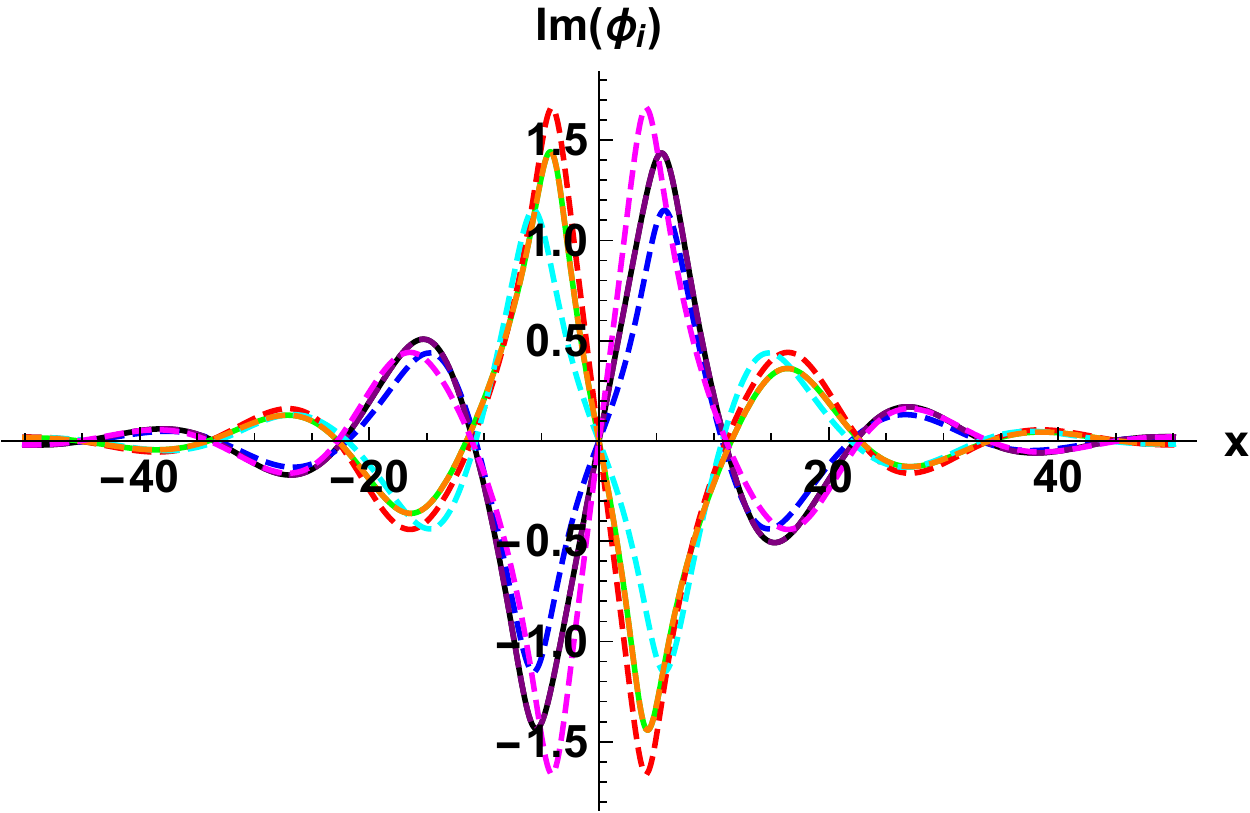}
     	\end{minipage}
     	\caption{Complex kink $\phi^k$ and anti-kink $\phi^a$ solutions to the self-dual and anti-self-dual BPS equations for the complex coupled sine-Gordon potential (\ref{potential in 1+1 example}). As initial conditions we have taken either $\phi(0) = \pi/2$ or $\phi(0) = -\pi/2$, as can be identified in panel (a), and the coupling constants are chosen as $\lambda=3$ and $\mu=1/2$.   }
     	\label{fig: Numerical plot of kink solutions}
     \end{figure}
     
  These solutions connect different types of vacua, which are easily identified as the fixed points of the dynamical system described by the self-dual and anti-self-dual BPS equations. We find the infinite set of vacua
    \begin{eqnarray}
    	v_1^{(n,m)} = \left(\text{arcsin}\mu + 2n\pi ,~m\pi \right)~,~~~v_2^{(n,m)} = \left(\pi-\text{arcsin}\mu + 2n\pi ,~m\pi \right),
    \end{eqnarray}
    with $n,m \in \mathbb{Z}$. Classifying the nature of the fixed points using standard techniques for dynamical systems by applying the linearization theorem, we can identify the nature of the fixed points as stable or unstable and conclude \cite{FringBPS} that the solutions only interpolate between the unstable vacua $v_1^{(n,2m)}$ and stable vacua $v_2^{(n,2m+1)}$ as indicated in figure \ref{fig: Stream line of kink}.  
    \begin{figure}[h]
    	\centering
    	\begin{minipage}[b]{0.5\textwidth}           \includegraphics[width=\textwidth]{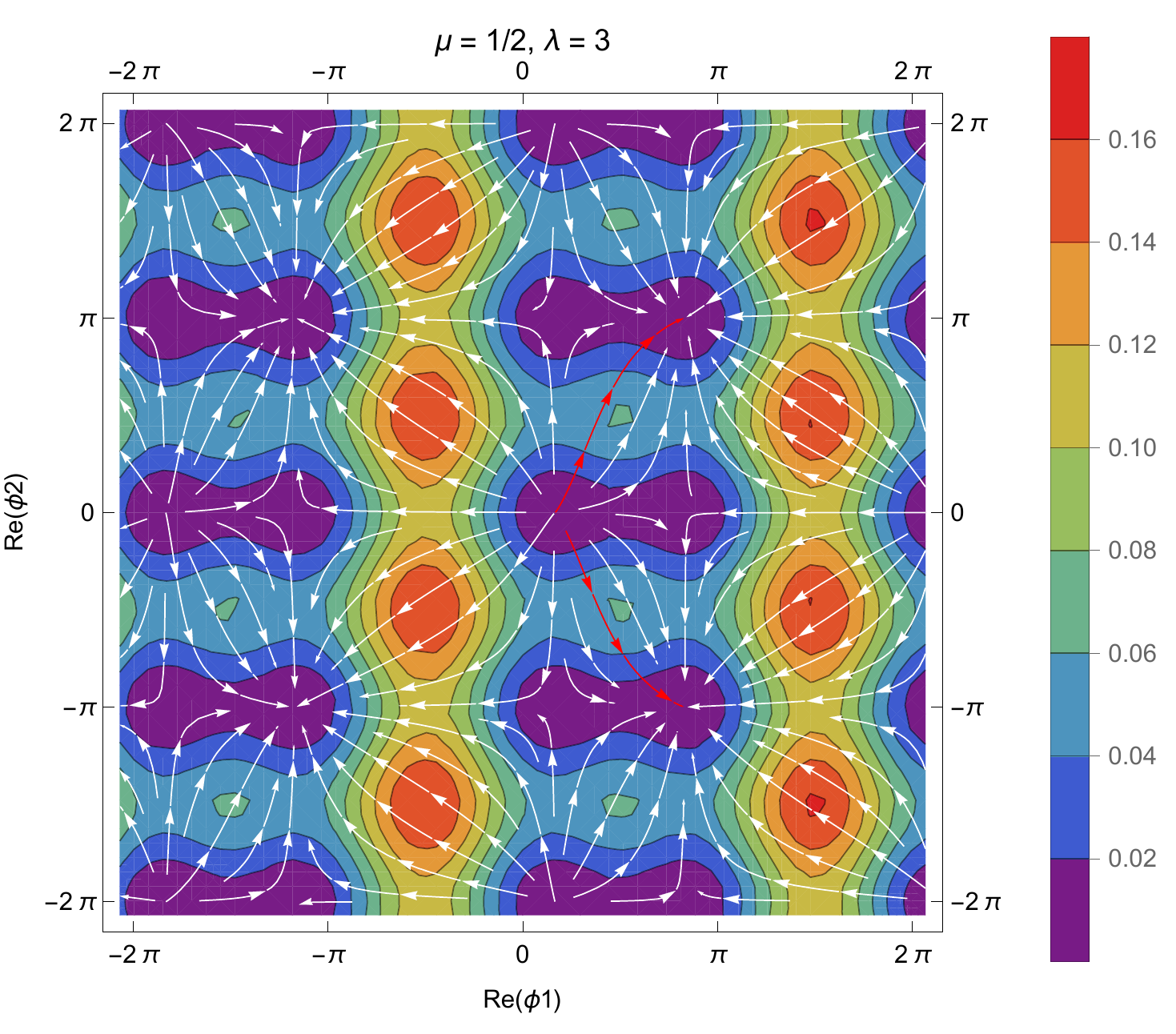}
    	\end{minipage}
    	\caption{Real part of the complex coupled sine-Gordon potential (\ref{potential in 1+1 example}) as a functions of the real parts of the fields $\phi_1$ and $\phi_2$ with real parts of the gradient flow superimposed. }
    	\label{fig: Stream line of kink}
    \end{figure}
    
    More specifically, the solutions depicted in figure \ref{fig: Numerical plot of kink solutions} interpolate the following vacua
    \begin{equation}
    	v_1^{(0,0)}\xrightarrow{\phi_1^{k+}\phi_2^{k+}}v_2^{(0,1)},~v_1^{(0,0)}\xrightarrow{\phi_1^{a+}\phi_2^{a+}}v_2^{(-1,1)},~v_1^{(0,0)}\xleftarrow{\phi_1^{a-}\phi_2^{k-}}v_2^{(0,-1)},~v_1^{(0,0)}\xleftarrow{\phi_1^{k-}\phi_2^{a-}}v_2^{(-1,1)}.
    \end{equation}
    The interpolation for other vacua with higher values for $n,m$ can be obtained by including the $n$ dependencies into the solutions.

    Having characterized the behaviour of these solutions, we evaluate their energies by means of (\ref{E1}). In the Hermitian case we can use the analytic expressions from (\ref{solution 2 of coupled sine-Gordon model}) obtaining
    \begin{equation}
    	E=2\left[1+\sqrt{1-\mu^2}-\mu~ \text{arctan} \left(\frac{\sqrt{1-\mu^2}}{\mu}\right)\right]. \label{topE}
    \end{equation}
   Since the energy is of a topological nature and the asymptotic behaviour is independent of the coupling constant, we obtain for $|\mu|<1$ the same real energy (\ref{topE}) also in the non-Hermitian case for all values of $\lambda$.     
    
        Finally let us use the symmetry argument (i)-(iii) to explain why the energies for these solutions must be real. The symmetry condition on the Hamiltonian (\ref{condition i}) is satisfied by the modified $\mathcal{CPT}$ symmetry 
        \begin{equation}\label{CPT symmetry of 1+1 example}
            \mathcal{CPT}:  \phi_i (x) \rightarrow (-1)^{\delta_{i2}} \left[\phi_i (-x)\right]^\dagger   .
        \end{equation}
        On the level of the equations of motion this symmetry maps the self-dual BPS equations to the conjugate of the anti-self-dual BPS equations. Thus the symmetry inevitably relates two different types of solutions. Inspecting the numerical solutions, we can indeed identify a set of self-dual and anti-self-dual solutions that are related by this $\mathcal{CPT}$-symmetry 
        \begin{equation}\label{symmetries of kink solutions}
            \phi_1^{k\pm} (x) = \left[\phi_1^{a\mp}(-x)\right]^\dagger ~,~~ \phi_2^{k+}=-\left[\phi_2^{k-}(-x)\right]^\dagger~,~~~\phi_2^{a+}=-\left[\phi_2^{a-}(-x)\right]^\dagger.
        \end{equation}
        Using now the fact that the energies are independent of the $\pm$-signs that relate the self-dual and the anti-self-dual equation and moreover that kinks and anti-kinks have identical energies we have guaranteed that the energies of the field related by the $\mathcal{CPT}$-symmetry are degenerate. Hence condition (\ref{condition iii}) is also satisfied, and therefore the energies related to these solutions must be real.
        
        More examples involving different types of complex potentials, such as a complex extended sine-Gordon model or a logarithmic potential that admits super-exponential kink solutions can be found in \cite{FringBPS}.

\subsection{Dual BPS theories in 3+1 dimensions, the BPS Skyrme model}
    In \cite{FringSkyrmion} we extended the study of complex BPS solutions to $3+1$ dimensions for several variants the original Skyrme model \cite{skyrme1962unified} that was originally designed as a theory of meson fields that also contains fermionic states interacting with those mesons. Nowadays it is interpreted as a low-energy effective field theory for quantum chromodynamics. Here we only present a summary of the main results obtained for a particular BPS version proposed in \cite{adam2010skyrme}. Similarly as demonstrated in the previous two sections, also this version allows for elegant analytic solutions, but in addition it resolved various discrepancies between the experimental results and the theoretical description by means of the original non-BPS full Skyrme model. The solutions to the latter are known to give binding energies that disagrees with the experiment \cite{full_skyrme_compared_with_experiment}.
    
    The model we consider here is a complex BPS Skyrme Lagrangian of the form
    \begin{equation}\label{complex Skyrmion Lagrangian}
    	\mathcal{L}_{\text{cBPSS}} = -\frac{\lambda^2}{4} \left(\sin\zeta - i\epsilon \cos\zeta\right)^4 \sin^2 \Theta \mathcal{B}_\mu \mathcal{B}^\mu -\mu^2 \left(\sqrt{1-\epsilon^2}-\cos\zeta - i\epsilon\sin\zeta\right),
    \end{equation}
where $\mathcal{B}^{\mu }:=\varepsilon ^{\mu \nu \rho \tau }\zeta _{\nu
}\Theta _{\rho }\Phi _{\tau }$. Redefining the coupling constants as $\lambda \rightarrow \tilde{\lambda}%
=\lambda (1-\epsilon ^{2})$, $\mu \rightarrow \tilde{\mu}=\mu (1-\epsilon
^{2})^{1/4}$, the model can be thought of as a complex boosted version of the 
BPS Skyrme model $\mathcal{L}_\text{BPSS}:=-\Tilde{\lambda}^2 N_0^2 B_\mu B^\mu~ -\Tilde{\mu}^2 V$ with potential $V=\frac{1}{2}\limfunc{Tr}\left( \mathbb{I-}U\right) $ originally proposed in \cite{adam2010skyrme}. The Skyrme fields $\zeta, \Theta, \Phi$ enter through a parametrization of a $SU(2)$-group valued elements $U$, which can be used to define a Lie algebraic current $L_{\mu }$ in form of a right Maurer Cartan form, which in turn may be used to define the topological current $B_{\mu }$ as
\begin{equation}
U:=e^{i\zeta (\sigma \cdot \vec{n})},~~~    L_{\mu }:=U^{\dagger }\partial _{\mu }U,~~~B^\mu := \frac{1}{N_0 }\epsilon^{\mu\nu\rho\tau} Tr\left(L_\nu L_\rho L_\tau\right) =\frac{1}{2N_{0}}\sin ^{2}\zeta \sin \Theta \,\mathcal{B}^{\mu}.
\end{equation}%
Here $\sigma$ denotes standard Pauli matrices, we take the three component unit vector as $\Vec{n}= (\sin\Theta \cos\Phi , \sin\Theta \sin \Phi , \cos\Theta)$ and $\lambda, \mu, \epsilon \in \mathbb{R}$ are constant parameters. $N_{0}$ is a normalization constant which is usually taken to
be $N_{0}=24\pi ^{2}$ for static solutions in order to produce integer Baryon numbers.

We notice that both Lagrangians $\mathcal{L}_{\text{cBPSS}}$ and $\mathcal{L}_{\text{BPSS}}$ are invariant under the anti-linear $\mathcal{CPT}$ transformation $\zeta \rightarrow -\zeta , i \rightarrow -i$. Moreover, their associated Hamiltonians are related by the adjoint action
 of a modified version of a Dyson map previously used in \cite{Bender:2005hf,FringBPS}
    \begin{equation}
        \eta = \exp\left[-\text{arctanh }\epsilon \int d^3 x \Pi^{\zeta} (t,r) \right] ,
    \end{equation}
     as $\mathcal{H}_{\text{cBPSS}}= \eta^{-1} \mathcal{H}_{\text{BPSS}} \eta$. The effect of this map is to shift the field $\zeta$ by $-i \text{arctanh} \epsilon$. 
    
    The real Lagrangian is known to posses a topological static solution referred to as a BPS Skyrmion found in \cite{adam2010skyrme} with the following compacton structure 
    \begin{equation}
        \zeta (r) = \left\{\begin{array}{ll}
            2\text{arccos}\left(\frac{1}{\sqrt{2}}\left|\frac{\Tilde{\mu}}{n\Tilde{\mu}}\right|^{1/3}r\right) \qquad & \text{for} \,\, r\in \left[0,r_c = \sqrt{2}\left|\frac{\Tilde{\mu}}{n\Tilde{\mu}}\right|^{1/3}r\right] \\
            0  & \text{otherwise}
        \end{array}\right. , 
    \end{equation}
where the Skyrmion fields are related to the spherical space-time coordinates $(r,\theta ,\phi )$,
$r\in \lbrack 0,\infty )$, $\theta \in \lbrack 0,\pi )$, $\phi \in
\lbrack 0,2\pi )$ as  $\Theta = \theta$, $\Phi = n\phi$ with  $n\in \mathbb{Z}$.

   The energy for these solutions was evaluated to be real and finite
    \begin{equation}
        E = \frac{64}{15} \sqrt{2} |n|\Tilde{\mu}\Tilde{\lambda} \pi (1-\epsilon^2)^{5/4}.
    \end{equation}
   Considering \cite{FringSkyrmion} instead the complex Lagrangian $\mathcal{L}_{\text{cBPSS}}$ in (\ref{complex Skyrmion Lagrangian} allows for a plethora of solutions, as we now briefly recall.
  We start by simplifying the calculation by re-casting the Lagrangian into the generic form (\ref{General form of the BPS energy}) with 
    \begin{eqnarray}
        A= \frac{\lambda}{2} (\sin\zeta - i \epsilon \cos \zeta)^2 \sin\Theta \mathcal{B}_0 ~,~~~\Tilde{A} = \mu \left(\sqrt{1-\epsilon^2}-\cos\zeta - i \epsilon \sin \zeta\right)^{1/2}.
    \end{eqnarray}
    Keeping the same identification between the Skyrme fields $\zeta, \Theta, \Phi$ and the spherical coordinates the BPS equations acquires the form
    \begin{equation}
        \frac{n\Tilde{\lambda}}{2 r^2} \sin^2 \left(\zeta - i \text{arctanh}\epsilon\right)\frac{d\zeta }{d r} = \pm \Tilde{\mu} \sqrt{1-\cos\left(\zeta- i \text{arctanh}\epsilon\right)}. \label{BPS1}
    \end{equation}
    We found an infinite set of solutions to this equation 
    \begin{equation}
        \zeta^\pm_{\alpha,m}(r) =\Tilde{\zeta}^\pm_{\alpha,m}(r) + i \text{arctanh}\epsilon= 2\text{arccos}\left[\omega^{\alpha} \frac{(n\Tilde{\lambda}c\mp \Tilde{\mu}r^3)^{1/3} }{\sqrt{2}n^{1/3}\Tilde{\lambda}^{1/3}}\right] + i \text{arctanh}\epsilon,
    \end{equation}
    with $\omega = \exp(i 2\pi /3)$, $\alpha \in \{0,1,2\}$, $m \in \mathbb{Z}$ and $c$ denoting an integration constant. We may view this equation as a complex shifted version of the solution to the real BPS equation with different boundary conditions, thus allowing for more possibilities of solutions not considered before. In particular, we would like to identify different types of compacton solutions. For this purpose we identify the critical values $r_0$ where $\Tilde{\zeta}^\pm_m (r_0)=0$ and $r_\pi$ where $\Tilde{\zeta}^\pm_m (r_\pi)=\pi$ as
    \begin{equation}
        r_{0,\alpha}^\pm := \omega^\alpha \left(\frac{\pm n \Tilde{\lambda}(c-2\sqrt{2})}{\Tilde{\mu}}\right)^{1/3}~,~~~ r_{\pi,\alpha}^\pm := \omega^\alpha \left(\frac{\pm n \Tilde{\lambda}}{\Tilde{\mu}}\right)^{1/3}.
    \end{equation}
    Besides recovering the standard BPS Skyrmion, we also find new types of solutions by pasting different branches together including a purely imaginary solution
     
    \begin{eqnarray}
        \zeta_\text{BPS} := \left\{\begin{array}{ccc}
            \Tilde{\zeta}^-_{0,0} &\text{ for }&0\leq r \leq r_0^-  \\
             0&\text{ for }&r_0^- < r
        \end{array}\right. ,&~~~\zeta_\text{St} := \left\{\begin{array}{ccc}
            \Tilde{\zeta}^-_{1,0} &\text{ for }&0\leq r \leq r_\pi^-  \\
            \Tilde{\zeta}^-_{0,0} &\text{ for }&r_\pi^-\leq r \leq r_0^-  \\
             0&\text{ for }&r_0^- < r
        \end{array}\right. ,\\
        \zeta_{i\text{BPS}} := \left\{\begin{array}{ccc}
            \Tilde{\zeta}^+_{0,0} &\text{ for }&0\leq r \leq r_0^+  \\
             0&\text{ for }&r_0^+ < r
        \end{array}\right.  . & \label{sol12}
    \end{eqnarray}
    Noting further that $r^+_\pi (c)=r^-_\pi (-c)$ we may even glue self-dual and anti-self dual solutions at this critical value obtaining 
    \begin{eqnarray}
        \zeta_\text{Cusp} := \left\{\begin{array}{ccc}
            \Tilde{\zeta}^+_{0,0} &\text{ for }&0\leq r \leq r_\pi^+ = r_\pi^-  \\
            \Tilde{\zeta}^-_{0,0} &\text{ for }&r_\pi^-\leq r \leq r_0^-  \\
             0&\text{ for }&r_0^- < r
        \end{array}\right. ,~\zeta_\text{Shell} := \left\{\begin{array}{ccc}
            0&\text{ for }&0\leq r \leq r_0^+  \\
            \Tilde{\zeta}^+_{0,0} &\text{ for }&r_0^+\leq r \leq r_\pi^+ =r_\pi^-  \\
            \Tilde{\zeta}^-_{0,0} &\text{ for }&r_\pi^-\leq r \leq r_0^-  \\
             0&\text{ for }&r_0^- < r
        \end{array}\right. . \quad  \label{sol23}
    \end{eqnarray}

 \begin{figure}[h]
	\centering
	\begin{minipage}[b]{0.49\textwidth}           \includegraphics[width=\textwidth]{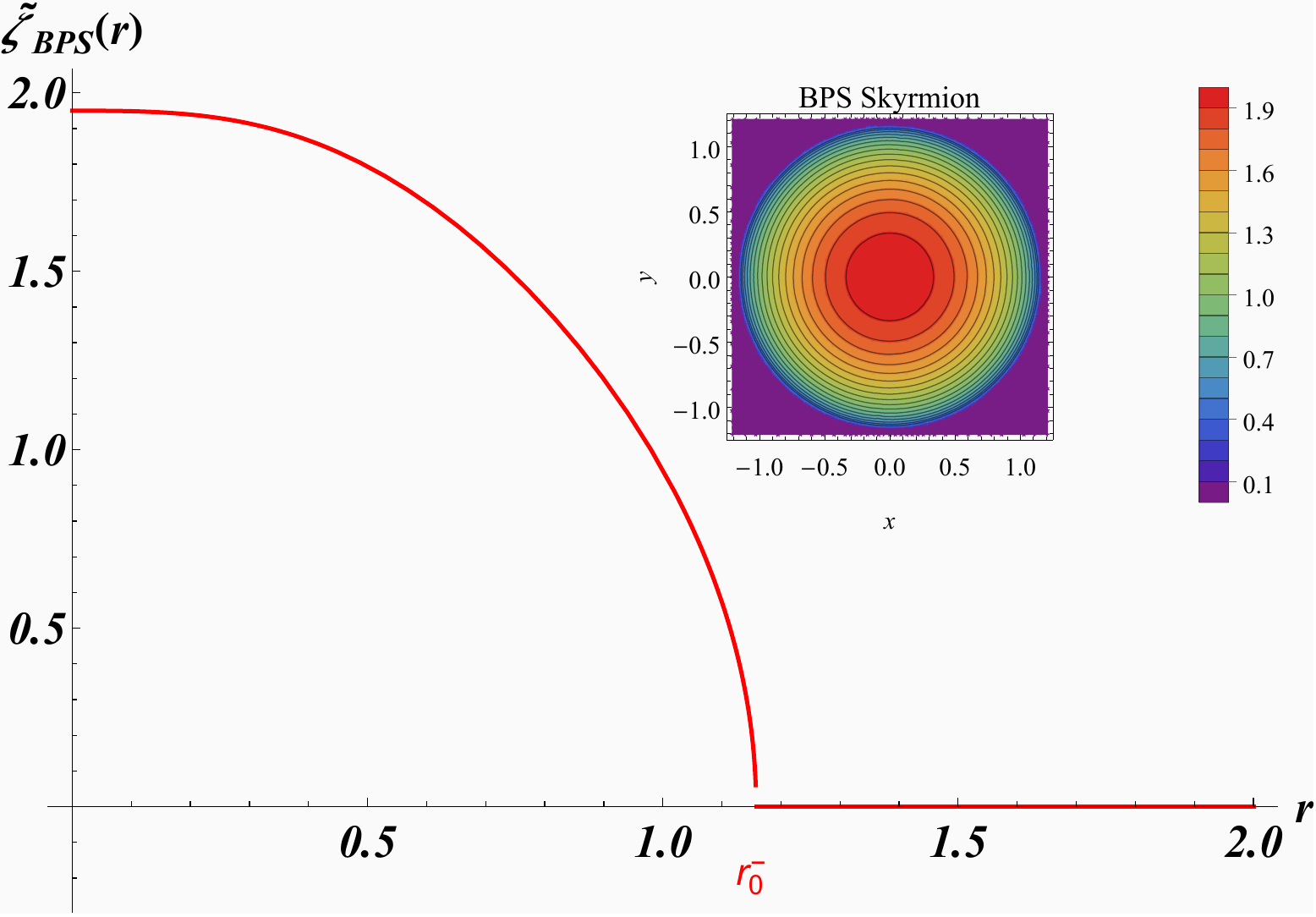} 
	\end{minipage}
\begin{minipage}[b]{0.49\textwidth}     
	\includegraphics[width=\textwidth]{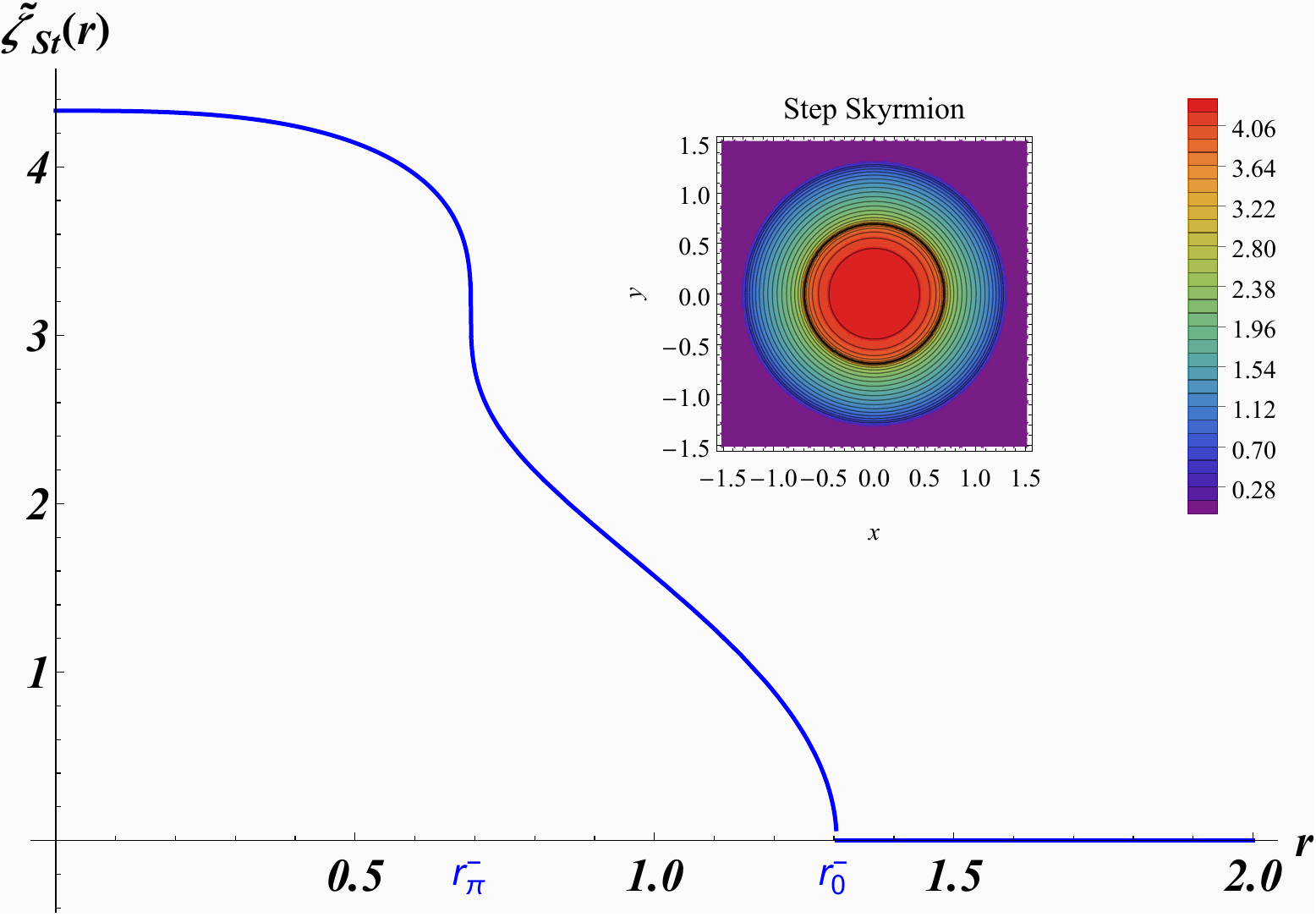}
\end{minipage}
	\begin{minipage}[b]{0.49\textwidth}           \includegraphics[width=\textwidth]{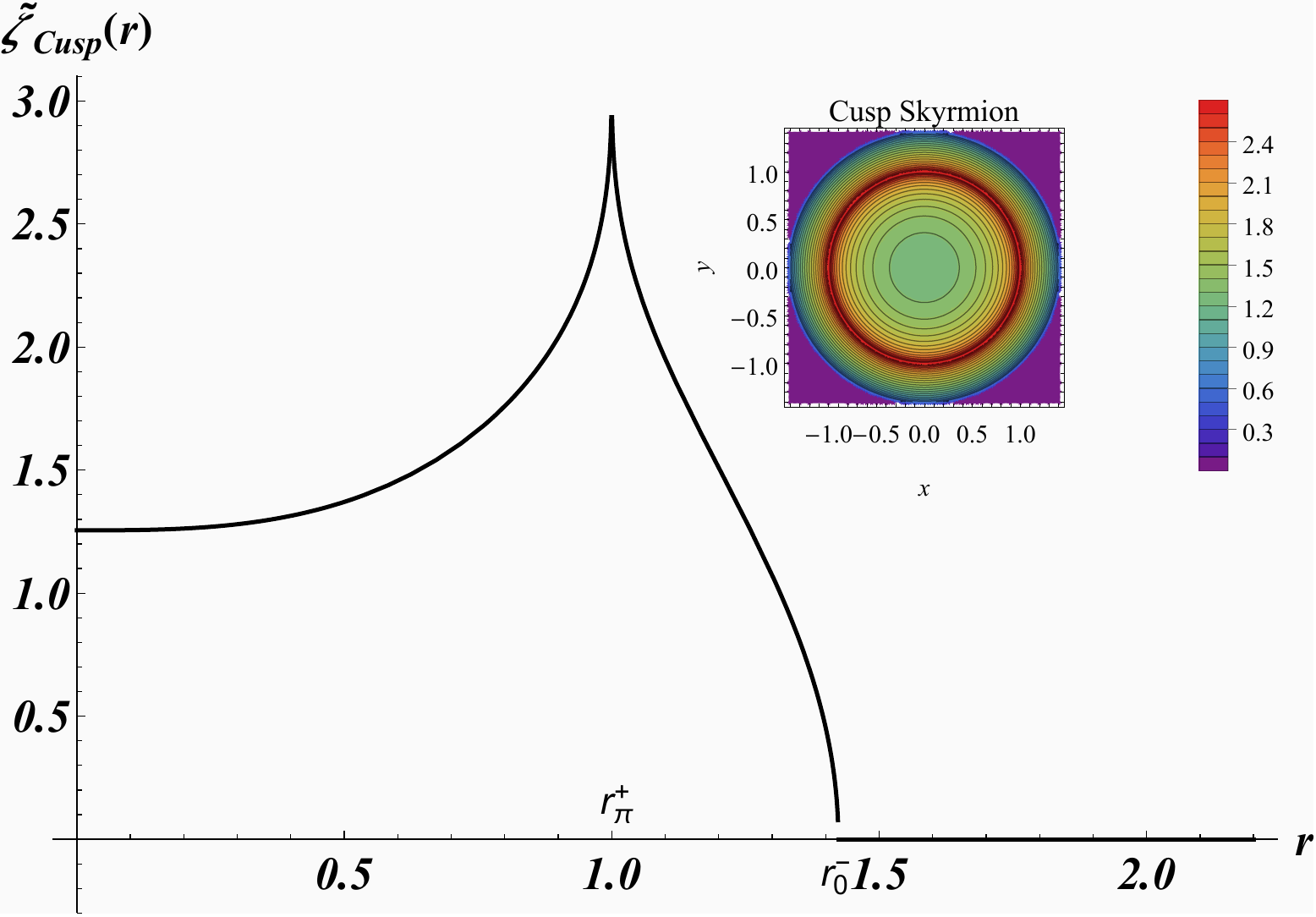}
	\end{minipage}
	\begin{minipage}[b]{0.49\textwidth}           
	\includegraphics[width=\textwidth]{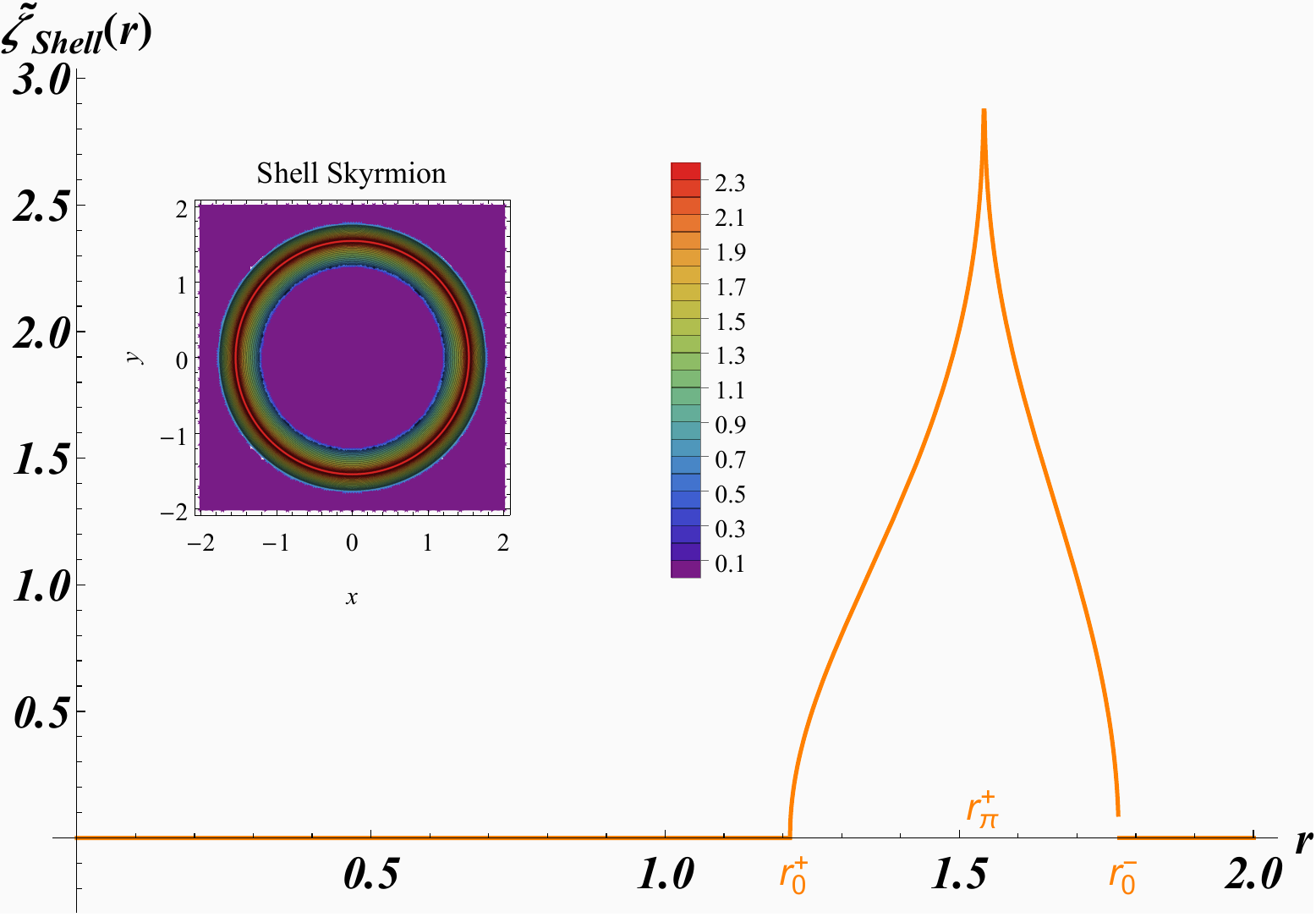}
\end{minipage}
	\caption{Differen types of Skyrmion solutions (\ref{sol12}) and (\ref{sol23}) of the equation of motion (\ref{BPS1})$^{2}$. We have have taken  $n=1$, $\tilde{\lambda} =3/2$, $\tilde{\mu} =2$, $s=1$ and $c=1/2$ (panel a), $c=-1/2$ (panel b), $c=-1$ (panel
		c), $c=1$ (panel d).}
	\label{comp solutions}
\end{figure}

We illustrate the behaviour of these solutions in figure \ref{comp solutions}. 

    The apparent issue that the first order derivative $d\zeta / dr$ is discontinues at the critical values is well known to occur also in the real case for the compacton solutions. We may argue therefore in a similar way as in \cite{adam2010skyrme} and conclude that the solutions are in fact well-defined. This is because the derivative always occurs with a multiple of $\sin^2 \zeta$ in the BPS equation so that the left and right limit of $\sin^2 \zeta d\zeta / dr$ have always a finite value. We note that this finite value differ in sign for Cusp and Shell solutions. This is because we have pasted the solutions from self and anti-self dual equations. However, this sign ambiguity disappears when one considers the equations of motion instead of the BPS equations. Hence, we adopt here the view that the equation of motion is more fundamental than the BPS equations and that all the solutions presented are therefore well-defined.
    
    The energies are computed to be finite and real, despite the solutions being complex
     \begin{equation}
    	E_{\text{BPS}/\text{St, Cusp}}=\frac{8}{15} n \Tilde{\mu}\Tilde{\lambda}\pi \left(8\sqrt{2}\mp 10 c \pm 3 c^{5/3}\right) ,~~~E_\text{Shell} = \frac{128}{15}\sqrt{2} n \Tilde{\mu}\Tilde{\lambda}\pi ,~~~
    	E_{i\text{BPS}}=-E_{\text{BPS}}.
    \end{equation}
    
    We conclude this section by verifying the reality conditions (i)-(iii). First we notice that condition (\ref{condition i}) is satisfied for our non-Hermitian Hamiltonian following from $\mathcal{L}_{\text{cBPSS}}$ by the anti-linear symmetry transformation 
    \begin{equation}
        \mathcal{CPT}' : \zeta (x_\mu ) \rightarrow \zeta^* (-x_\mu)+2 i \text{arctanh}\epsilon .
    \end{equation}
    Performing this transformation on our solutions, we find
    \begin{equation}
        \zeta^\pm_{\alpha,m} (r) \rightarrow \left[\zeta^\pm_{\alpha,m} (r)\right]^* +2i \text{arctanh}\epsilon = \zeta^\pm_{\alpha,m}(r).
    \end{equation}
    Thus in this case the solutions are mapped to itself so the condition (\ref{condition iii}) is automatically satisfied and the energy must therefore be real.
    
      In \cite{FringSkyrmion} we gradually make the above system more complicated and consider examples for different types of potentials and also different terms in the Lagrangians. In there we consider a model that gives rise to semi-kink and massless solutions, Bender-Boettcher type potentials, a complex trigonometric model that breaks the $\mathcal{CPT}$ symmetry and a new submodel of the BPS system. In all these examples the reality argument is shown to hold, albeit in different ways.

\section{Conclusion}\label{Section: conclusion}
    In a strictly pseudo-Hermitian approach, by utilizing Dyson maps acting adjointly on a non-Hermitian field theoretic Hamiltonian, we have demonstrated that the Goldstone theorem holds and the Higgs mechanism are realised in the parameter regime in which the theory admits a modified $\mathcal{CPT}$-symmetry. This parameter regime is bounded by three different types of boundaries: (i) the standard exceptional points, at which we have to identify the Goldstone bosons technically in a modified way due the Jordan structure of the squared-mass matrix so that the Higgs mechanism is still in tact. (ii) the zero exceptional points at which the Goldstone boson can not be identified and can therefore also not be coupled to the gauge particle to give them mass and (iii) the trivial vacua for which the continuous symmetry is not broken and we do not expect Goldstone bosons to emerge in the first place that may be utilized in the Higgs mechanism. In the $\mathcal{CPT}$-broken regime not only these mechanism break down, but the non-Hermitian theory becomes entirely ill-defined.  
    
    In a non-Hermitian $SU(2)$-invariant local gauge theory we have identified complex variants of the t'Hooft-Polyakov monopoles, which have real energies in the $\mathcal{CPT}$-symmetric parameter regime. Interestingly in two disconnected regions the solutions do not respect this symmetry themselves, but instead the symmetry exchanges the solutions to the dual and the self-dual solutions to the BPS equations. As the energies of these solutions are degenerate, our reality conditions holds and the energies are guaranteed to be real. In the intermediate region the solutions do not respect the $\mathcal{CPT}$-symmetry and are therefore also not expected to have real energies. We computed the energies explicitly confirming the general assertions.
    
    For BPS theories in 1+1 dimensions, a complex coupled sine-Gordon model, and in 3+1 dimensions, BPS versions of the Skyrme model, we have found solitonic/kink/anti-kink type soltions with real energies governed by the anti-linear $\mathcal{CPT}$-symmetry in a similar fashion as the monopole solutions. The broadening of the solutions to the larger complex regime allowed for new types of configurations we refer to as step, cusp and shell Skyrmion solutions with further examples to be found in \cite{FringSkyrmion}.

\newif\ifabfull\abfulltrue


\end{document}